\documentclass[modern]{aastex61}
\usepackage{natbib}
\pdfoutput=1
\shorttitle{Precovery of TESS Single Transits}
\shortauthors{Yao et al.}

\begin{document}


\title{Precovery of TESS Single Transits with KELT}


\author{Xinyu Yao}
\affiliation{Department of Physics, Lehigh University, 16 Memorial Drive East, Bethlehem, PA 18015, USA}

\author{Joshua Pepper}
\affiliation{Department of Physics, Lehigh University, 16 Memorial Drive East, Bethlehem, PA 18015, USA}

\author{B. Scott Gaudi}
\affiliation{Department of Astronomy, The Ohio State University, Columbus, OH 43210, USA}

\author{Jonathan Labadie-Bartz}
\affiliation{Department of Physics, Lehigh University, 16 Memorial Drive East, Bethlehem, PA 18015, USA}
\affiliation{Department of Physics and Astronomy, University of Delaware, Newark, DE 19716, USA}

\author{Thomas G. Beatty}
\affiliation{Department of Astronomy \& Astrophysics, The Pennsylvania State University, 525 Davey Lab, University Park, PA 16802, USA}
\affiliation{Center for Exoplanets and Habitable Worlds, The Pennsylvania State University, 525 Davey Lab, University Park, PA 16802, USA}

\author{Knicole D. Col\'{o}n}
\affiliation{NASA Goddard Space Flight Center, Exoplanets and Stellar Astrophysics Laboratory (Code 667), Greenbelt, MD 20771, USA}

\author{David J. James}
\affiliation{Harvard-Smithsonian Center for Astrophysics, 60 Garden St, Cambridge, MA 02138, USA}

\author{Rudolf B. Kuhn}
\affiliation{South African Astronomical Observatory, P.O. Box 9, Observatory 7935, South Africa}
\affiliation{South African Large Telescope, P.O. Box 9, Observatory 7935, South Africa}

\author{Michael B. Lund}
\affiliation{Department of Physics and Astronomy, Vanderbilt University, 6301 Stevenson Center, Nashville, TN 37235, USA}

\author{Joseph E. Rodriguez}
\affiliation{Harvard-Smithsonian Center for Astrophysics, 60 Garden St, Cambridge, MA 02138, USA}

\author{Robert J. Siverd}
\affiliation{Las Cumbres Observatory Global Telescope Network, 6740 Cortona Dr., Suite 102, Santa Barbara, CA 93117, USA}

\author{Keivan G. Stassun}
\affiliation{Department of Physics and Astronomy, Vanderbilt University, 6301 Stevenson Center, Nashville, TN 37235, USA}
\affiliation{Department of Physics, Fisk University, 1000 17th Avenue North, Nashville, TN 37208, USA}

\author{Daniel J. Stevens}
\affiliation{Department of Astronomy, The Ohio State University, Columbus, OH 43210, USA}

\author{Steven Villanueva, Jr.}
\affiliation{Department of Astronomy, The Ohio State University, Columbus, OH 43210, USA}

\author{Daniel Bayliss}
\affiliation{Department of Physics, University of Warwick, Gibbet Hill Road, Coventry CV4 7AL, UK}

\date{\today}

\begin{abstract}

During the TESS prime mission, 74\% of the sky area will only have an observational baseline of 27 days. For planets with orbital periods longer than 13.5 days, TESS can only capture one or two transits, and the planet ephemerides will be difficult to determine from TESS data alone. Follow-up observations of transits of these candidates will require precise ephemerides.  We explore the use of existing ground-based wide-field photometric surveys to constrain the ephemerides of the TESS single-transit candidates, with a focus on the Kilodegree Extremely Little Telescope (KELT) survey.  We insert simulated TESS-detected single transits into KELT light curves, and evaluate how well their orbital periods can be recovered.  We find that KELT photometry can be used to confirm ephemerides with high accuracy for planets of Saturn size or larger with orbital periods as long as a year, and therefore span a wide range of planet equilibrium temperatures. In a large fraction of the sky we recover 30\% to 50\% of warm Jupiter systems (planet radius of 0.9 to 1.1 $R_J$ and $13.5 < P < 50$ days), 5\% to 20\% of temperate Jupiters ($50 < P < 300$ days), and 10\% to 30\% of warm Saturns (planet radius of 0.5 to 0.9 $R_J$ and $13.5 < P < 50$ days). The resulting ephemerides can be used for follow-up observations to confirm candidates as planets, eclipsing binaries, or other false positives, as well as to conduct detailed transit observations with facilities like JWST or HST.

\end{abstract}


\section{Introduction}

Of the $\sim$3,700 exoplanets discovered to date, $\sim$2,900 of them are known to transit their host star.  The vast majority of known transiting planets have come from two types of surveys.  The space-based missions CoRoT \citep{Auvergne:2009}, Kepler \citep{Borucki:2010,Koch:2010}, and the reborn K2 mission \citep[]{Howell:2014}, have discovered nearly 2,600 transiting planets, representing 90\% of all known such planets.  The other main source of known transiting planets are the ground-based transit surveys that use relatively wide field telescopes such as SuperWASP \citep{Street:2003}, HATNet \citep{Bakos:2004}, HATSouth \citep{Bakos:2013}, KELT \citep{Pepper:2007,Pepper:2012}, XO \citep{McCullough:2005}, TrES \citep{Alonso:2004}, and QES \citep{Alsubai:2013}.

Discovering large numbers of exoplanets can reveal the underlying demographics and frequencies of various types of planets \citep{Mayor:2011,Howard:2012}, and for more recent analysis see \citet{Fulton:2017,Johnson:2017,Petigura:2017a,Petigura:2017b,Weiss:2017}.  Transiting planets are especially valuable because the transit itself constrains the planet radius and orbital inclination, and together with radial velocity observations, it provides the mass and density of the planet.  Furthermore, for transiting planets it is possible to study the atmosphere of the planet via transmission or emission spectroscopy \citep[]{Charbonneau:2002,Charbonneau:2005,Deming:2005} and constrain the planet's eccentricity with secondary eclipses.  However, very precise and high SNR spectroscopic observations are required for studies of exoplanet atmospheres (as well as the radial velocity measurements used to infer the planetary mass), which can only be conducted for systems with relatively bright host stars.

\begin{figure}[t]
\begin{center}
\makebox[\textwidth][c]{\includegraphics[width=1.0\textwidth]{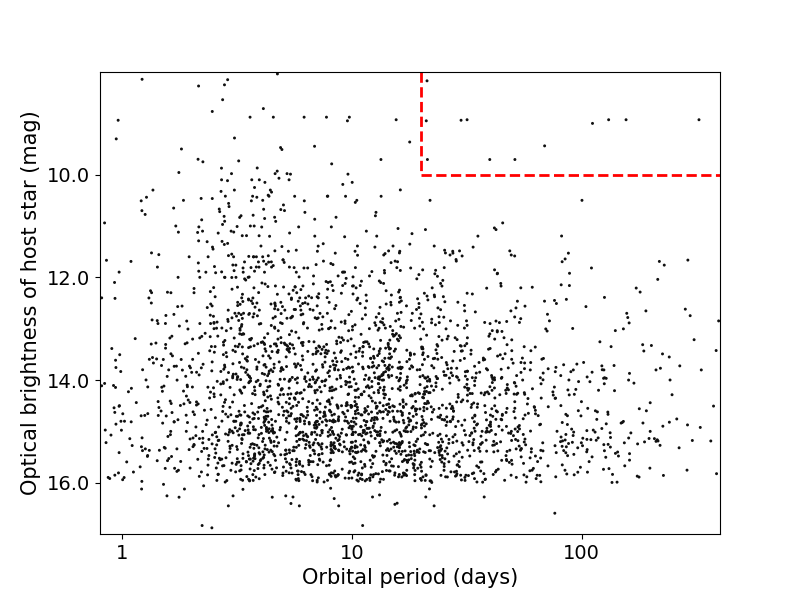}}
\caption{Optical magnitude versus orbital period of host stars of known transiting exoplanets. The region inside the red dashed lines shows the 12 exoplanets with bright hosts and orbital periods longer than 20 days.  The data displayed here comes from the NASA Exoplanet Archive, accessed on May 15, 2018.}
\label{fig:period-mag}
\end{center}
\end{figure}

The bulk of known transiting planets come from Kepler and K2, and most of those orbit relatively faint host stars. Of the 1776 transiting planet host stars with a known optical magnitude discovered by the space missions Kepler, K2, or CoRoT, only 85 have an optical magnitude brighter than $V<12$, and only 9 are brighter than $V<10$\footnote{https://exoplanetarchive.ipac.caltech.edu/}.  The ground-based surveys are generally sensitive to brighter host stars; 125 of the 224 host stars of planets discovered by those surveys are brighter than $V\sim 12$, and 23 are brighter than $V\sim 10$.  However, most of the planets discovered by the ground-based surveys are in very short-period orbits, with none of the ground-based discoveries having orbital periods longer than 20 days. Figure \ref{fig:period-mag} shows that only 12 known transiting planets have host stars brighter than $V\sim 10$ and also have orbital periods longer than 20 days. This is due to a well-known selection effect against longer period planets in transit surveys \citep[e.g.][]{Pepper:2003,Gaudi:2005}. Since atmospheric studies of planets benefit from being able to probe a range of planetary equilibrium temperatures, this region of parameter space (planets on long periods orbiting bright host stars) must be populated with new discoveries before a more comprehensive study of exoplanet atmospheres can be undertaken.

The Transiting Exoplanet Survey Satellite (TESS) is designed to detect transiting planets of bright stars across the whole sky.  However, most of the TESS transit detections will have short orbital periods, due to the short TESS observing time for most of the sky ($\sim27$ days).  Estimates of the yield of planets from TESS indicates that several hundred planets will be detected by TESS with only a single transit seen \citep{Villanueva:2018,Cooke:2018}. In this paper, we explore the potential of combining data from an existing ground-based survey (specifically, the Kilodegree Extremely Little Telescope [KELT] survey) with TESS observations.  The goal is to refine the orbital properties of the long-period transiting exoplanets orbiting bright hosts discovered by TESS, which will be extremely valuable for the ability to probe transiting planets at a range of equilibrium temperatures.  We test the feasibility of this endeavor by simulating single transit events as detected by TESS, and then inserting them as periodic signals into real KELT light curves.  We then use standard techniques to recover those signals from the KELT data, and characterize how the recovery rate depends on the properties of the simulated planets and stars.  The approach is effectively ``precovery" of TESS-detected signals from existing data, in which we leverage the information provided by the TESS single transit detections as constraints on the automated search for transits in the existing ground-based survey data. Our use of the term ``precovery" in this paper refers to the use of existing, archival observational data to derive information about a later-identified object or system, a term commonly used in the investigation of solar system bodies.

The process of searching for single transit signals in Kepler and K2 data has been described in several papers. \citet{Foreman-Mackey:2016} used a probabilistic model comparison method to search for single transits in Kepler data. That process conducted an initial candidate search using a box-shaped matched filter, and then checked candidate events using a comparison between a physical transit model and a set of known systematics to eliminate false positives. They discovered 16 systems with likely astrophysical transits or eclipses out of the brightest $\sim40,000$ G/K dwarfs in the Kepler target list.  \citet{Osborn:2016} used least-square minimization after removing long-term variability to fit pre-generated transit models to search for single transits in K2 data, finding such events in seven candidates.  None of the single transit candidates from these searches have yet been confirmed, although one has been statistically validated \citep{Beichman:2018}.

One of the key goals of TESS is to discover transiting planets orbiting bright stars to provide a rich set of targets for detailed atmospheric measurements by HST and JWST.  Spectroscopic observations can reveal a planet's atmospheric composition, structure, and dynamics, and is a key objective of JWST \citep[]{Beichman:2014}. Atmospheric observations using transmission or emission spectroscopy, verification of planet eccentricity and albedo through secondary eclipse observations, spin-orbit alignment measurements, and transit timing observations all require the ability to schedule observations at future transit times.  Given the high demand for space telescope time, transit observations must be known to a precision of roughly 30 minutes, otherwise precious observing time on high-demand telescopes will potentially be wasted.

It is already known that the ephemerides of even multiple-observed transits degrade over time, and any improvement offered by archival data can save such systems for future space-based transit observation.  \citet{Benneke:2017} demonstrated that this kind of ephemeris improvement was necessary for a 33-day period candidate (K2-18b) observed in the 80-day {\it K2} observing window (see also \citet{Steffanson:2018} for more examples).

In this paper, we explicitly do not attempt to model or predict the number of single-transit detections that TESS will make (see \citet{Villanueva:2018} and \citet{Cooke:2018} for such predictions), nor the number of those that the KELT data will be able to precover.  Rather, what we investigate here is the overall efficiency of the precovery procedure to establish the ephemerides of single-transit systems across a range of empirical properties, regardless of the abundance of such systems that are actually seen by TESS.

We discuss the nature of the data used to conduct this analysis in Section \ref{sec:data}, including a description of why the KELT survey is specifically useful for this project. Section \ref{sec:methods} describes the methodology for inserting transits and recovering the signals. Section \ref{sec:results} presents the  results of the recovery tests, describing the types of transiting planets for which this method can be successfully applied.  In Section \ref{sec:disc} we discuss the significance of our findings and ways to build upon these initial results.

\section{Data}
\label{sec:data}


\subsection{The TESS mission}

TESS \citep{Ricker:2015} is a NASA mission that will monitor bright and nearby stars for transiting planets, with a prime mission lasting two years.  The combined field of view is $24^{\circ} \times 96^{\circ}$ with four 4096$\times$4096 pixel CCDs, at $\sim$21\arcsec\ per pixel, and a bandpass between 600 nm and 1000 nm, referred to as $T$-band.  It will acquire observations in two modes.  A selection of about 200,000 stars will be observed at a 2-minute cadence, while all stars in the field of view will be observed at a 30-minute cadence, including about 30 million stars brighter than $T=15$.  The short-cadence observations are commonly referred to as the postage-stamp targets, while the longer cadence observations are referred to as Full Frame Image (FFI) targets.

The 2-min TESS targets are selected based on the suitability for detection of small transiting planets.  That requirement leads generally to a selection of bright, cool, dwarf stars.  The target selection process must account for a number of factors, and can lead to a range of expected planet yields.  Papers that have looked at the expected yields include \citet{Sullivan:2015}, \citet{Bouma:2017}, and \citet{Barclay:2018}.  Each of these papers simulates the TESS survey, and the resulting physical and observational properties of the detected planets.  They find that while the key goals of TESS generally center on small planet discovery, TESS will also discover hundreds of planets larger than 4$R_{\Earth}$ from the 2-min observations, and of order thousands such planets from the FFIs, with rough agreement between those predictions about the numbers of planets that will be detected of different sizes.  

TESS will observe the sky in a set of pointed observations in which the spacecraft will nearly continuously observe a section of the sky stretching from the ecliptic to the ecliptic pole for 27 days, with each section referred to as a sector.  The mission will step around ecliptic longitude over the course of 13 sectors to cover most of an ecliptic hemisphere over the course of a year, and then rotate and observe the other hemisphere.  Near the ecliptic poles, subsequent sectors will overlap, so that stars in those regions can be observed for many months.  Near the ecliptic poles, the overlapping sectors create the TESS continuous viewing zones (CVZs).  In those regions, many stars will have near-continuous coverage over 351 days (although, due to gaps between adjacent CCDs, most stars even in the CVZs will miss at least one sector of observation).  However, the CVZs cover only 1.7\% of the sky (2\% of the TESS footprint in the prime mission), and so the majority of the sky observed by TESS (74\%) will have an observational baseline of only $\sim$27 days. For those transiting exoplanets that have orbital periods longer than 13.5 days, TESS can only capture one or two transits, and for periods longer than 27 days, TESS can only capture one transit. In these cases, the true ephemerides of the planets will be difficult to determine from TESS data alone. \citet{Sullivan:2015} predict that long period planets (P$>$20d) will account for approximately 20\% of the TESS transit detections, while \citet{Bouma:2017} find similar expectations. 

Since follow-up confirmation observations are expensive in time and resources, the ability to determine the precise ephemerides for these systems is extremely valuable. Even though the orbital period can be roughly estimated from a single transit \citep{Seager:2003,Yee:2008,Osborn:2016}, additional information is still needed to precisely determine these ephemerides and reduce the associated uncertainty in the transit time.  These estimates require an assumption or constraint on the orbital eccentricity, as well as a precise estimate of the density of the host star, and are often fairly imprecise, thereby necessitating either extensive photometric follow-up, or radial velocity observations \citep{Yee:2008}.

\citet{Villanueva:2018} have done a comprehensive simulation and investigation into the likely TESS single-transit population.  They predict about 1200 single-transit events in TESS.  Those numbers are higher than that of the other simulation efforts by factors of 20\% to 50\%, but possibly because the other simulations are not as complete at the long-period end.  The actual number of single-transit detection will depend on many observational factors unknown at the present, but all indications are that there will a minimum of several hundred such cases.

\subsection{The KELT survey}

\begin{figure}[t]
\begin{center}
\includegraphics[width=1.0\textwidth]{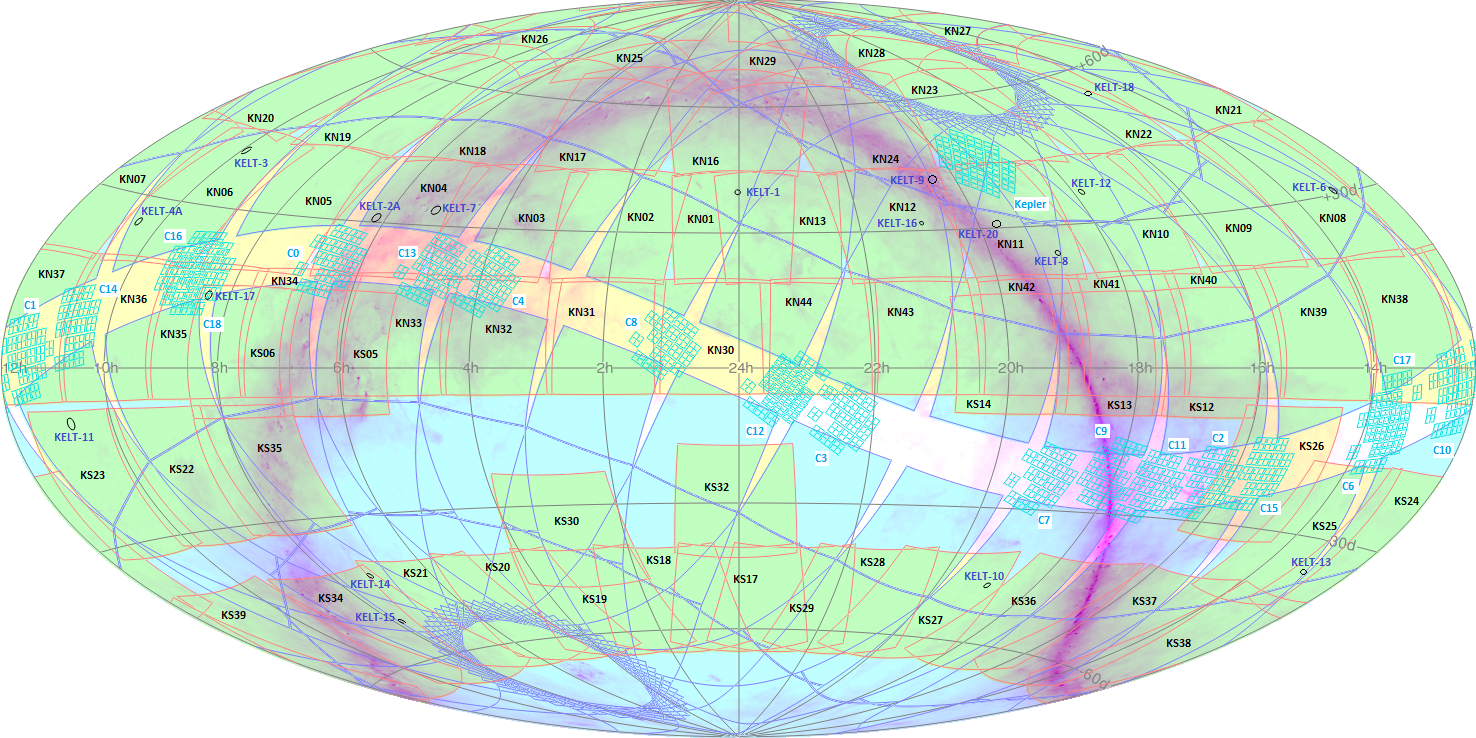}
\caption{Equatorial map of the sky, with the galactic plane displayed in pink. Colors indicate the locations of the KELT fields (green), the Kepler and K2 fields (cyan), and a sample location layout for the TESS fields (blue).}
\label{fig:fields}
\end{center}
\end{figure}

The Kilodegree Extremely Little Telescope (KELT) transit survey \citep{Pepper:2007,Pepper:2012} is a photometric survey using two small-aperture (42 mm), wide field ($26^{\circ} \times 26^{\circ}$) telescopes located at Winer Observatory in Arizona, US, (KELT-North) and the SAAO observing station in Sutherland, South Africa, (KELT-South). Each camera contains a CCD with 4096$\times$4096 pixels and provide a pixel scale of roughly 23\arcsec\ pixel$^{-1}$. The effective passband that KELT uses is roughly equivalent to a broad $R$-band filter. The KELT fields cover about 70\% of the sky including a very large fraction of the TESS footprint in both year 1 and year 2 of the TESS mission.  Figure \ref{fig:fields} shows a map of KELT fields, along with Kepler and K2 fields, and a sample set of TESS fields.  Table \ref{tab:typ} lists the names of KELT fields with the sky coordinates of the field centers. Fields are observed every season since their first observation, with observations starting between 2005 and 2013.  Although some fields have been observed for over twelve years, the reduced data available for the analysis in this paper only contained observations for the first eight years of the survey, which sets the maximum time baseline used in the analysis here.  The observing procedure leads to an average cadence of about 20 minutes, with $\sim$1500 observations every year.  All KELT fields overlap at least partially with TESS fields. The photometric precision of KELT ranges from several mmag RMS at the bright end of its magnitude range, around $V=7.5$, where stars begin to saturate in normal KELT observations, to $\sim$5\% RMS around $V=13$. The KELT survey is optimized for the discovery of transiting planets in the range $8<V<10$, and to date has discovered 23 exoplanets with transit depths ranging from 3 to 14 mmag (see \citet{Pepper:2018} for references).


KELT has several features which make it especially useful for combination with TESS.  The magnitude range of greatest photometric sensitivity overlaps with that of TESS.  The wide field of KELT ($26^{\circ} \times 26^{\circ}$) allows it to cover a large fraction of the entire sky, and the relatively stable and regular observing strategy provides light curves with consistent noise and sampling properties over long time baselines.

Though the KELT fields also overlap with the Kepler field and some K2 fields, the approach described in this paper is unlikely to be useful for those data sets.  Most stars observed by Kepler and K2 are too faint to have high-quality light curves when observed by KELT. In addition, the observational baseline for Kepler is four years, and for the K2 mission most fields have time baseline of 80 days, compared to 27 days for TESS.  Because of the smaller number of stars with precision light curves, and the fact that the transit precovery procedure is more efficient for shorter periods, the potential value of this approach is likely very small for application to Kepler/K2 data. 

\begin{deluxetable}{ccc|ccc|ccc}

\tablecaption{KELT field locations (J2000.0)}
\label{tab:typ}
\tablehead{\colhead{Field} & \colhead{RA (deg)} & \colhead{DEC (deg)} & \colhead{Field} & \colhead{RA (deg)} & \colhead{DEC (deg)} & \colhead{Field} & \colhead{RA (deg)} & \colhead{DEC (deg)}}
\startdata
        N01  &  001.50  & 31.67 & N21  &  200.80  & 57.00 & S19  & 046.00  & -53.00 \\
        N02  &  030.52  & 31.67 & N22  &  240.80  & 57.00 & S20  & 069.00  & -53.00 \\
        N03  &  059.54  & 31.67 & N23  &  280.80  & 57.00 & S21  & 091.80  & -53.00 \\
        N04  &  088.56  & 31.67 & N24  &  320.80  & 57.00 & S22  & 138.00  & -20.00 \\
        N05  &  117.58  & 31.67 & N25  &  054.09  & 79.00 & S23  & 161.00  & -20.00 \\
        N06  &  146.60  & 31.67 & N26  &  126.46  & 79.00 & S24  & 184.00  & -30.00 \\
        N07  &  175.62  & 31.67 & N27  &  197.61  & 79.00 & S25  & 207.00  & -30.00 \\
        N08  &  204.64  & 31.67 & N28  &  269.76  & 79.00 & S26  & 230.00  & -20.00 \\
        N09  &  233.66  & 31.67 & N29  &  342.12  & 79.00 & S27  & 299.00  & -53.00 \\
        N10  &  262.68  & 31.67 & S05  &  091.80  & 03.00 & S28  & 322.00  & -53.00 \\
        N11  &  291.70  & 31.67 & S06/N14  & 114.90  & 03.00 & S29  & 345.00  & -53.00 \\
        N12  &  320.72  & 31.67 & S12/N15  & 253.00  & 03.00 & S30  & 045.00  & -36.00 \\
        N13  &  349.74  & 31.67 & S13  & 276.00  & 03.00 & S35  & 115.05  & -20.00 \\
        N16  &  000.80  & 57.00 & S14  & 299.00  & 03.00 & S36  & 261.00  & -53.00 \\
        N17  &  040.80  & 57.00 & S15  & 322.00  & 03.00 & S37  & 226.80  & -53.00 \\
        N18  &  080.80  & 57.00 & S16  & 345.00  & 03.00 & S38  & 192.60  & -53.00 \\
        N19  &  120.80  & 57.00 & S17  & 000.00  & -53.00 & S39  & 158.40  & -53.00 \\
        N20  &  160.80  & 57.00 & S18  & 023.00  & -53.00 &   &   &   \\ 
\enddata
\end{deluxetable}

\section{Methods}
\label{sec:methods}

\subsection{General Approach}

Although the KELT survey has a lower photometric precision and duty cycle than TESS, the KELT fields cover a large fraction of the TESS footprint in both the first year of TESS observations of the southern ecliptic hemisphere and the second year of TESS observations of the ecliptic north (Figure \ref{fig:fields}), and the KELT light curves have long time baselines. Furthermore, KELT data are essentially uncorrelated on timescales of longer than a few hours.  When phased and binned on 20-hour timescales (the duration of a long-period transiting planet, the RMS of the detrended KELT lightcurves is typically 0.5 mmag. Thus, for single-transit TESS detections with deep enough transits, KELT can detect the signals in phase-folded data, and determine the ephemerides of the TESS single-transit candidates.  Our overall approach is therefore to simulate TESS single-transit detections, insert the transits into a selection of KELT light curves, and attempt to recover the inserted signals using standard techniques.  

\subsection{Selection of KELT Light Curves}

In the analysis for this paper, we made use of KELT light curves already obtained for the survey transit search (see \citet{Siverd:2012} and \citep{Kuhn:2016} for details).  Since the reduction and extraction of KELT light curves is a time-intensive process, we did not make use of the entire set of KELT data, but we will use the full KELT data set by the time TESS light curves are publicly released.

KELT uses a German equatorial mount, which requires a “flip” as it tracks stars past the meridian. Thus, the optics and detector are rotated 180 deg with respect to the stars between observations in the Eastern and Western sides of the meridian. As a result of that meridian flip, a given star experiences different detector defects, optical distortions, PSF shape, flat-fielding errors, etc., in each orientation. Those conditions require us to treat observations in the east and west essentially as two separate data sets \citep{Siverd:2012}.  We also use the Trend Filtering Algorithm (TFA) \citep[]{Kovacs:2005} to correct various types of systematic noise.  This results in 4 separate light curves for each KELT star: an uncorrected east version, and uncorrected west version, and a TFA-processed version of each.

We first select high quality KELT light curves, meaning those with relatively low RMS.  In each KELT field, we create a plot of the RMS of each KELT light curve (before and after TFA-processing) versus approximate V-band magnitude, bin the data in magnitude space between $V=9$ and $V=11$, and fit a curve to the 70th percentile of RMS in each bin.  We define the set of light curves below that fitted curve and also below a fixed limit of 30 mmag RMS as the ``high-quality" KELT light curves for that field.  We also require that each light curve must have at least 3000 observations.  This process is carried out separately for all the KELT light curves in each field orientation.

\begin{figure}[t]
\begin{center}
\makebox[\textwidth][c]{\includegraphics[width=1.0\textwidth]{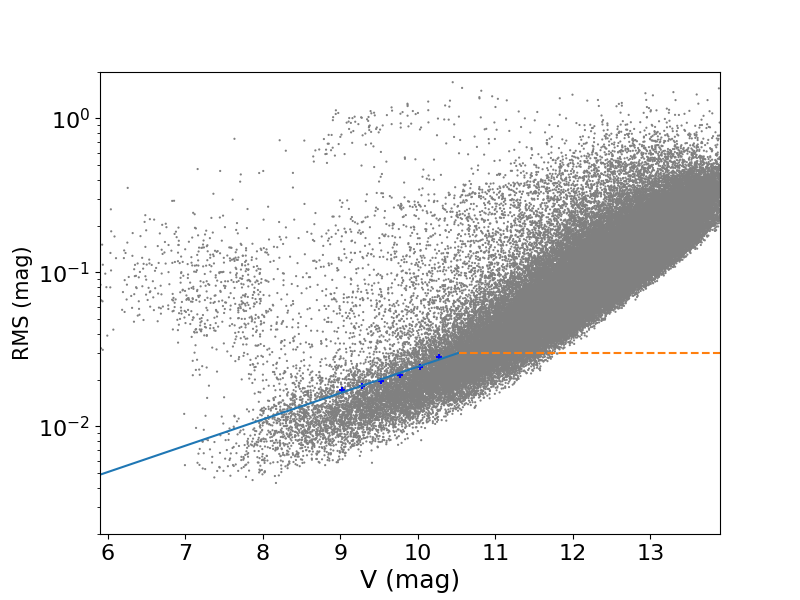}}
\caption{RMS vs. approximate V magnitude for raw data in one example field (west field of N05) with selection criteria. The orange dashed line represents the 30 mmag cut; blue dots correspond to the 70th percentile RMS in each 0.25 magnitude bin between V=9 and V=10.25, and the blue solid line represents the fit to the dots.  Stars brighter than $V\sim7.5$ experience saturation and nonlinearity effects, and so have large scatter.}
\label{fig:RMS}
\end{center}
\end{figure}

An example of the selection method in one field is shown in Figure \ref{fig:RMS}.  The orange line represents the 30 mmag limit of RMS, while the blue line is fitted to the 70th percentile of the RMS range in each bin. Stars whose light curves (in both raw and detrended version) fall below both the orange and blue lines are selected for analysis.

If both the east and west light curve of a star pass these cuts, then the light curves are combined into a single light curve and used for later steps. 
In all, $\sim$340,000 stars are selected from the 4.5 million KELT stars with light curves.  The selected stars are then cross-matched with the TESS Candidate Target List \citep[CTL;][]{Stassun:2017}.  The CTL contains stellar information such as TESS mag (T-mag), $T_{eff}$, stellar mass, and radius.

We end up with $\sim$133,000 combined KELT light curves matched to the TESS CTL ($\sim$70,000 light curves in northern KELT fields and $\sim$63,000 light curves in southern KELT fields).  These light curves are primarily distinguished from the rest of the KELT light curves by their photometric precision, and as such have similar numbers of data points and time baseline as the rest of the KELT data, since there are few light curves in the full data set that have fewer than 3000 total data points.
The fact that less than half the high-quality KELT light curves match to the CTL is not surprising.  The CTL contains only stars that are especially useful for transit detection, meaning that giant stars and very hot stars (OBA spectral types) are generally excluded, while no such restriction was applied to the KELT light curves.

\begin{figure}[t]
\begin{center}
\makebox[\textwidth][c]{\includegraphics[width=1.2\textwidth]{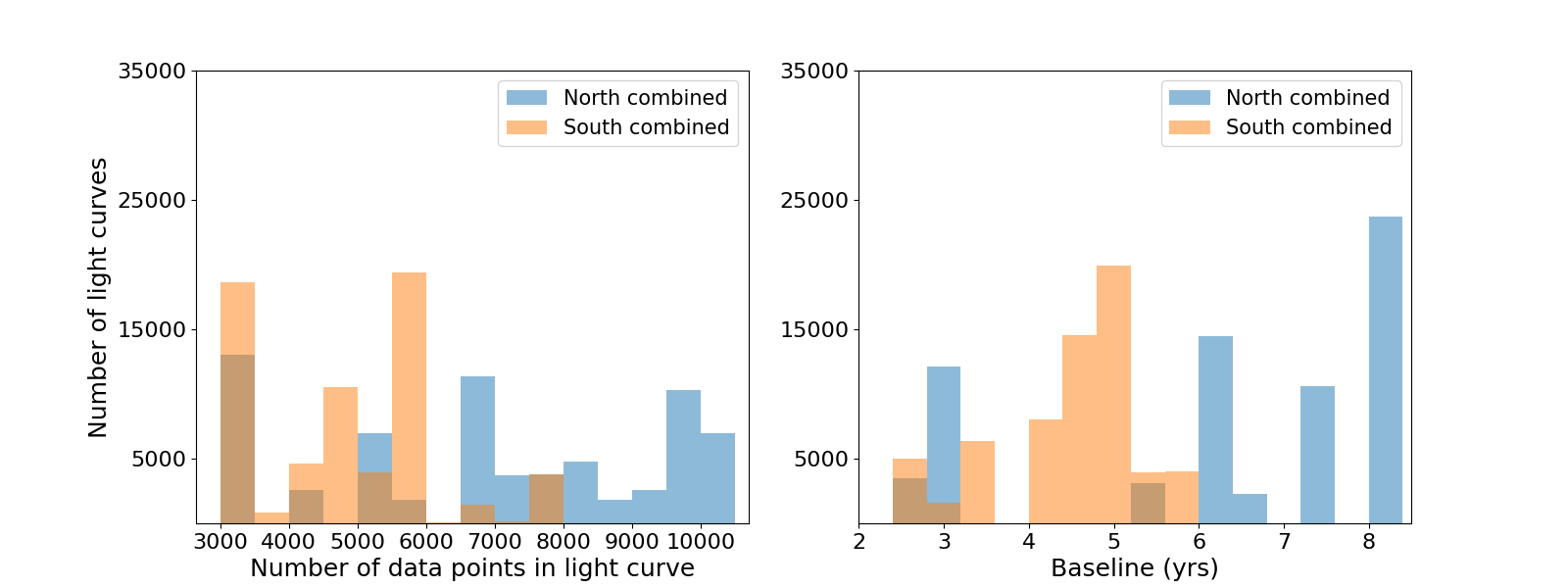}}
\caption{Histogram distributions of the number of data points (left panel) and time baseline (right panel) of matched KELT-North and KELT-South combined light curves used in this analysis.}
\label{fig:hist}
\end{center}
\end{figure}

We display distributions of the number of observations and observational time baseline for all KELT light curves that pass these cuts in Figure \ref{fig:hist}.  Because KELT-North has been observing for longer than KELT-South, and generally has more observations and a longer time baseline, we split the two sets of light curves.  For the selected KELT-North light curves, the number of observations is irregularly distributed between 3,000 and 11,000, while the baselines range from 2.5 years up to $\sim$8 years. For the selected KELT-South light curves, the number of observations is irregularly distributed between 3,000 and 8,000, while the baselines range from 2.5 years up to $\sim$6 years. The KELT survey is ongoing, and the number of observations and length of the observing baseline will continue to grow as new observations are made, and as more existing data are reduced.

\begin{figure}[t]
\begin{center}
\makebox[\textwidth][c]{\includegraphics[width=1.2\textwidth]{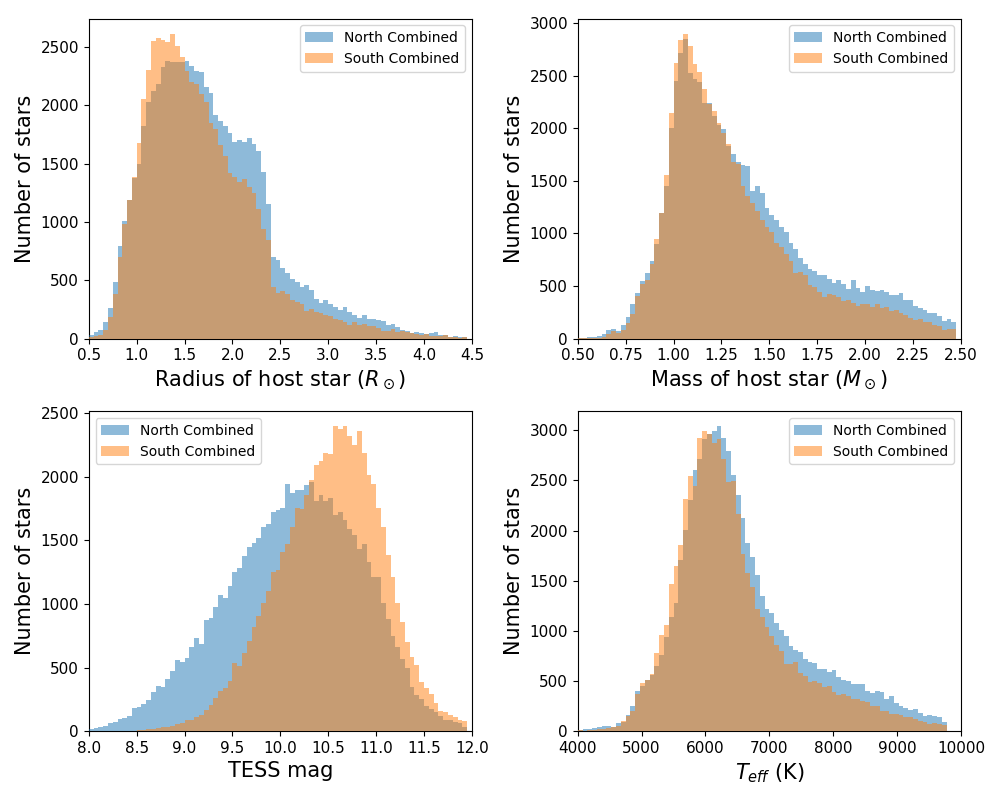}}
\caption{Histograms of stellar properties (radius, mass, TESS magnitude and effective temperature) for the matched KELT-North and KELT-South combined light curves}.
\label{fig:stellar-overall}
\end{center}
\end{figure}

KELT is a magnitude-limited survey.  As such, it is affected by selection biases including Malmquist bias \citep{Bieryla:2015}.  As a result of that, plus the selection of the brighter, low-RMS light curves, the stars that make it past the cuts tend to be solar mass or greater.  Figure \ref{fig:stellar-overall} illustrates the stellar properties for the matched KELT-North and KELT-South combined light curves.  Most stars are generally F-type, with radii between 1 and 2.5 $R_{\odot}$, masses between 1 and 2 $M_{\odot}$, and effective temperatures between 5500 and 7500 K.  The TESS magnitudes range mostly from $9 < T < 11$.

\subsection{Inserting Simulated Transits}
\label{sec:inserting}

We assume that the transit depth $\delta$, transit duration $T$, ingress/egress duration $\tau$, and mid-transit times $T_C$
will be easily measurable in TESS data for the events relevant to this work. Since the smallest RMS for the brightest stars in KELT data is a few mmag, we are only considering events detected in TESS with depths of this order and larger. This corresponds very roughly to planetary radii larger than $4 R_{\Earth}$ for planets transiting mid-type main sequence stars.

Our overall approach is to insert transit signals into KELT light curves, and to then search for those signals based on the expected TESS light curve information.  Since we are trying to determine the transit detection sensitivity of this approach, we begin with a range of simulated transit properties.  We generate $T_C$ randomly between July 1st, 2018 and July 1st, 2020 corresponding to the expected start and end of the primary TESS mission, the orbital period was assigned randomly from 13.5 days to 300 days in log space, and the transit depth was assigned randomly from 3 mmag to 20 mmag in log space.  

We use a Mandel-Agol \citep{Mandel:2002} transit model to simulate the transits.  We assume equatorial transits in a circular orbit ($b=0$).  However, since that model requires physical parameters, we use the stellar mass and radius from the TIC, the $T_C$ and Period as defined above, and calculate the planetary radius that corresponds to the desired transit depth.  We must also incorporate limb-darkening to calculate transit depth.  We adopt quadratic limb darkening coefficients for the KELT bandpasses from \citet{Claret:2012,Claret:2013} (The limb darkening coefficients in Johnson-Cousins R band were adopted).

The particular ranges of transit depth and orbital period we probe here are based on a rough accounting of where the KELT data will be most useful. The single-transit events seen by TESS will generally have minimum orbital periods of half the 27-day observing baseline for a sector, or 13.5 days.  Although most TESS transit detections at the shorter end of the 13.5 to 27 day range will in fact display two transit in TESS, our goal in this paper is to investigate the value of KELT data for those without multiple transits.  We use 300 days as a convenient outer range for periods to probe, although see the discussion at the end of the paper for comments on that limit. We selected a minimum depth of 3 mmag since that is about the shallowest transit for which a transit can be detected in KELT data, and a maximum depth of 20 mmag since that is approximately the greatest transit depth observed to date, although there are a handful of known transiting planets with larger transit depths. We also reject those cases where the transit signal corresponds to a planet with a radius larger than twice Jupiter's radius ($\sim$15\% of the total sample).

\begin{figure}[t]
\begin{center}
\makebox[\textwidth][c]{\includegraphics[width=1.2\textwidth]{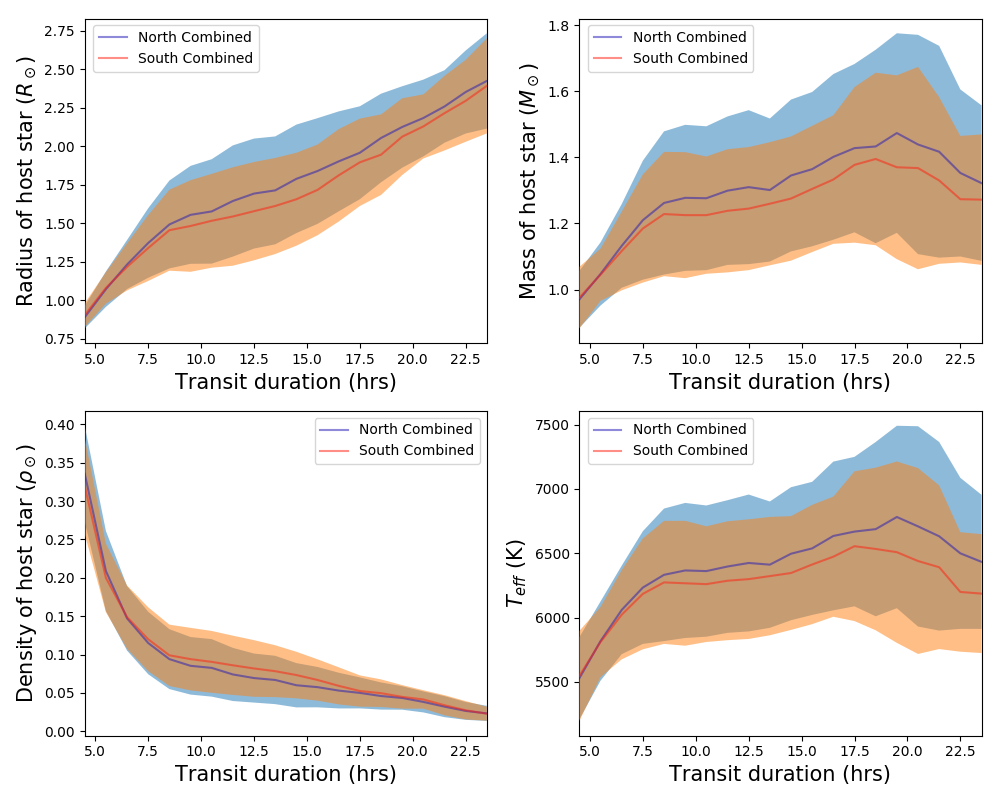}}
\caption{Distributions of stellar properties (radius, mass, density and effective temperature) in transit duration parameter space for the matched KELT-North and KELT-South combined light curves, with stellar properties taken from the TIC. The shaded color region corresponds to the median absolute deviation in one hour bins.  Note that since the TIC determines stellar mass based on $T_{eff}$ (see section 3.3.2 of \citet{Stassun:2017}), the plot of stellar mass looks almost identical to that of temperature.}
\label{fig:stellar-depth-duration}
\end{center}
\end{figure}

Since transit signals are inserted randomly into the light curves, the properties of the host stars are uniform across the range of transit depths and orbital periods.  However, they are not uniform across the physical parameters of the stars.  We can calculate the theoretical transit duration using a combination of equations (18) and (19) of \citet*{Winn:2010}:
\begin{equation}
\label{eqn:duration}
T = 13\;\mathrm{hr}
\left(\frac{P}{365\;\mathrm{days}}\right)^{1/3}\left(\frac{\rho_{\star}}{\rho_{\odot}}\right)^{-1/3}.
\end{equation}
The stellar density $\rho_{\star}$ is calculated from stellar mass and radius obtained from CTL-7. We can then relate the inserted orbital periods, along with the stellar densities, to plot the distribution of stellar properties across transit duration in Figure \ref{fig:stellar-depth-duration}.  We see that the distribution of stellar parameters is clearly not uniform across transit duration. Specifically, smaller, higher density, cooler stars exhibit shorter transit durations. Since the transit duration for a given orbital period depends on the mass and radius of the star, signals that were inserted with a given period will have shorter duration for smaller stars.  Long duration transits generally correspond to those with long orbital periods, as in equation \ref{eqn:duration}. But they also scale with the low stellar densities and hence large stellar radii, and are therefore preferentially hotter stars or sub-giants. These features have important implications for the use of this analysis for TESS users, as explained in the discussion below.
 
\subsection{Recovery of Inserted Transits}
\label{sec:recovery}

We use the box-fitting least-squares algorithm \citep[(BLS);][]{Kovacs:2002}, as implemented in the VARTOOLS \citep{Hartman:2016} software package, to search for transits in the KELT data.  We use the version of BLS with a fixed transit duration, a fixed depth, and a fixed ingress/egress duration at a given $T_C$. The concept behind this approach is that since the parameters $T_C$, $T$, $\delta$, and $\tau$ for a given transit signal will be known with high fidelity from the TESS observations, we can then search the KELT light curves for repeating transit signals with the same transit time, duration, depth, and $\tau$ as seen in the TESS data.  

In order to conduct that search, we need to know the transit duration, $\tau$, and depth as TESS would observe.  We therefore simulate the transit signal just as in \S \ref{sec:inserting} using the Period, the stellar mass and radius, and planet radius as a continuous, noiseless light curve, and measure the duration of the transit. This time, we use the limb-darkening coefficients appropriate for the TESS bandpass from \citet{Claret:2017}.  We then calculate $\tau$ by subtracting the theoretical transit duration from equation \ref{eqn:duration} above from the duration determined from the modeled TESS light curve.

We searched the KELT light curves for signals with periods ranging from 13.5 days to 300 days, with a frequency resolution of 300,000.  We fix the $T_C$ based on the originally generated value for that target, and fix the transit duration, depth, and $\tau$ based on the TESS light curve simulation.

It should be reiterated that the selection of ``high-quality" KELT light curves is based only on RMS at a given magnitude.  Beyond the RMS cuts, specific types of variability are not investigated or excluded, such as astrophysical variability due to rotation, pulsation, or eclipses.  In fact, it is quite likely that real transits exist currently in the KELT data, not just the inserted transits.  We do not expect such signals to interfere with the recovery of the inserted signals.  That is because we use a BLS search with a fixed $T_C$ and transit duration, which means that only signals that fit those parameters are evaluated.  The probability that any real transits in the data happen to match the randomly chosen $T_C$ and duration parameters is insignificant, and so should not affect the overall results.




\section{Results} 
\label{sec:results}

For the bulk of the analysis we present here, we investigate only the collective results using KELT-North lightcurves.  Compared to KELT-South, the KELT-North data cover longer time baselines and have more total observations, and will generally yield larger rates of precovery.  We do that partly to provide an optimistic perspective on the value of this effort for TESS.  Also, since the data available at the time of this analysis does not include the last one to four years of KELT data (still being reduced), the larger current KELT-North data set is a better reflection of the KELT data available for precovery in the next year.  For completeness sake, we include summary plots of the KELT-South precovery rates at the end of the paper.

\begin{figure}[t]
\begin{center}
\makebox[\textwidth][c]{\includegraphics[width=1.0\textwidth]{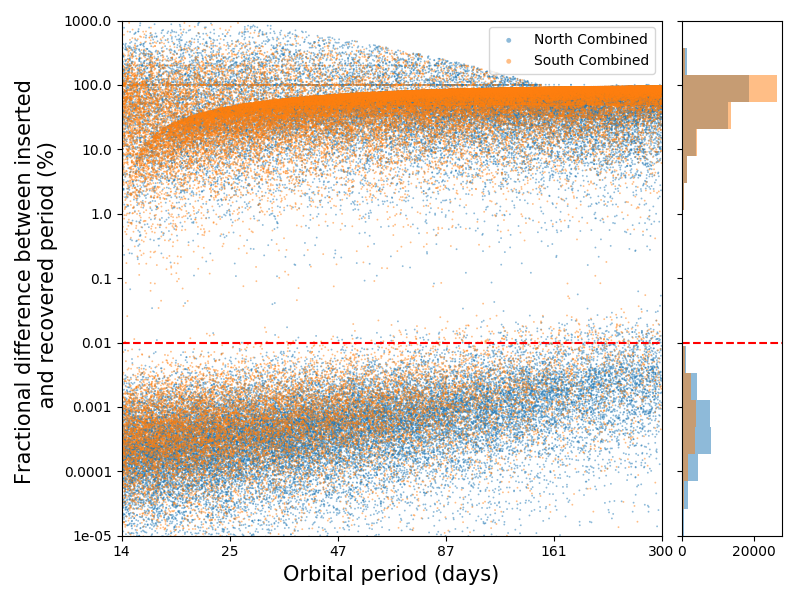}}
\caption{The histogram and scatter distributions of the fractional period recovery with inserted orbital period for KELT North and KELT South light curves. The red dashed lines represent 0.01\%.} 
\label{fig:period_percent}
\end{center}
\end{figure}

For a successful recovery of an inserted transit signal, we require that the percent error between the output period and the inserted period be within 0.01\%.  That limit was selected because we find it to naturally divide the populations of recovered versus failed cases, as shown in Figure \ref{fig:period_percent}.  The exact value of the cut is somewhat arbitrary, but can be shifted with little effect on the overall results.  However, for the frequency spacing used in this analysis (300,000), the cut cannot be much smaller than that, without running into the resolution of the retrieved period at the long-period end. 

This criterion cannot be used to determine actual recovery of signals with real data, since it is not an observable, but we can use it to gauge the accuracy of the precovery process for different star, planet, and light curve properties.  To examine the practical use of the precovery approach, we investigate the use of the SNR of the recovered transit below in section \ref{sec:prac}.

As is common in searches for periodic variability, we find that a small number of the searches identify a period that is a small fractional multiple of the true period, at a rate much higher than random chance, but not as high as the true period recovery rate.  Overall, we find that between 1\% and 7\% of the inserted signals are found within 0.01\% of either 1/2, 3/2, or twice the inserted period.  In the analysis below, we do not include those cases as successful recoveries, but we discuss how they might be used in the discussion section.

We can examine how well our results match theoretical expectations. The SNR of a transit \citep{Pepper:2003,Gaudi:2005} is:
\begin{equation}
\label{eqn:SNR}
SNR = \sqrt{N f_{tr} }\frac{\delta}{\sigma}
\end{equation}
where $N$ is the total number of data points in light curve, $f_{tr}$ is the fraction of time the planet is in transit ($T/P$), $\delta$ is the transit depth and $\sigma$ is the fractional photometric uncertainty.  We can express SNR in terms of the stellar density and orbital period:
\begin{equation}
\label{eqn:SNR_period}
SNR = \sqrt{N} \frac{\delta}{\sigma} {\left(\frac{3}{G\pi^2}\right)}^{1/6}P^{-1/3}\rho_{\star}^{-1/6}
\end{equation}
We can also express SNR in terms of transit duration:
\begin{equation}
\label{eqn:SNR_duration}
SNR = \sqrt{N} \frac{\delta}{\sigma} {\left(\frac{3}{G\pi^2}\right)}^{1/2}T^{-1}\rho_{\star}^{-1/2}
\end{equation}

From these equations, we can see that the SNR is proportional to the transit depth, while it has negative correlation with stellar density, orbital period, and transit duration.

\subsection{Recovery results}

\begin{figure}[t]
\begin{center}
\makebox[\textwidth][c]{\includegraphics[width=1.0\textwidth]{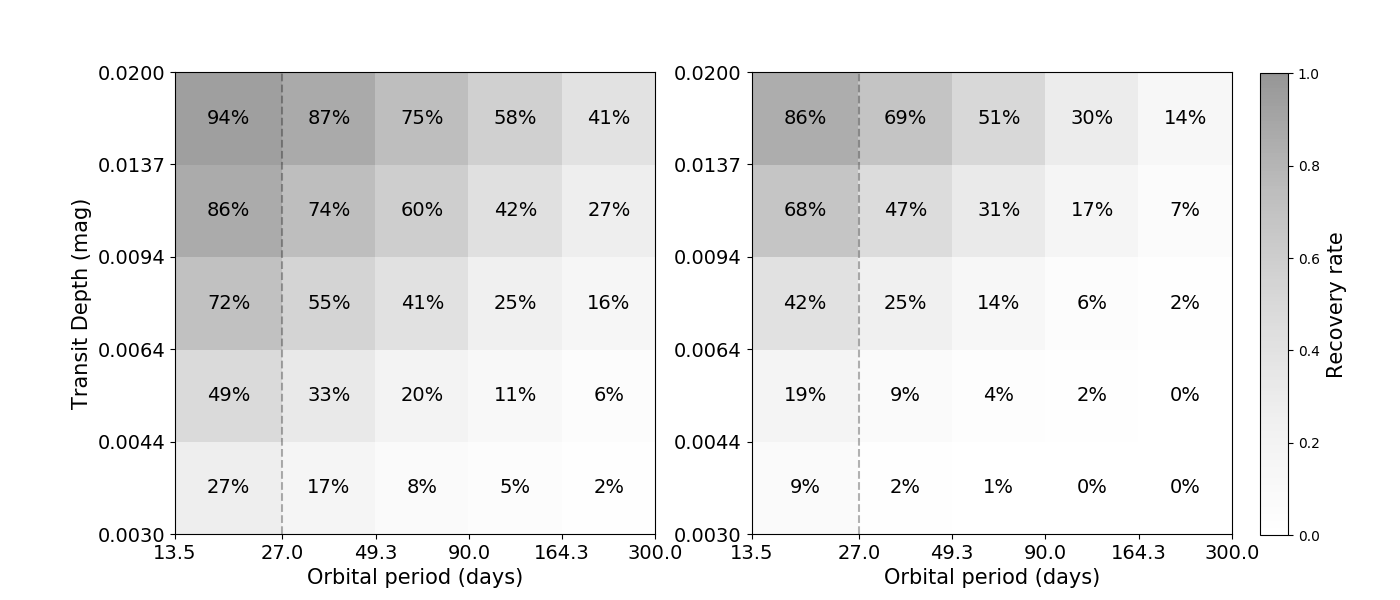}}
\caption{Recovery rates of simulated transits for orbital period-transit depth bins for the KELT-North (left) and KELT-South (right) light curves. The grayscale bar indicates the fraction of the transits that are correctly recovered, which is also represented by the percent value in each bin.} 
\label{fig:rate_period}
\end{center}
\end{figure}

\begin{figure}[t]
\begin{center}
\makebox[\textwidth][c]{\includegraphics[width=1.0\textwidth]{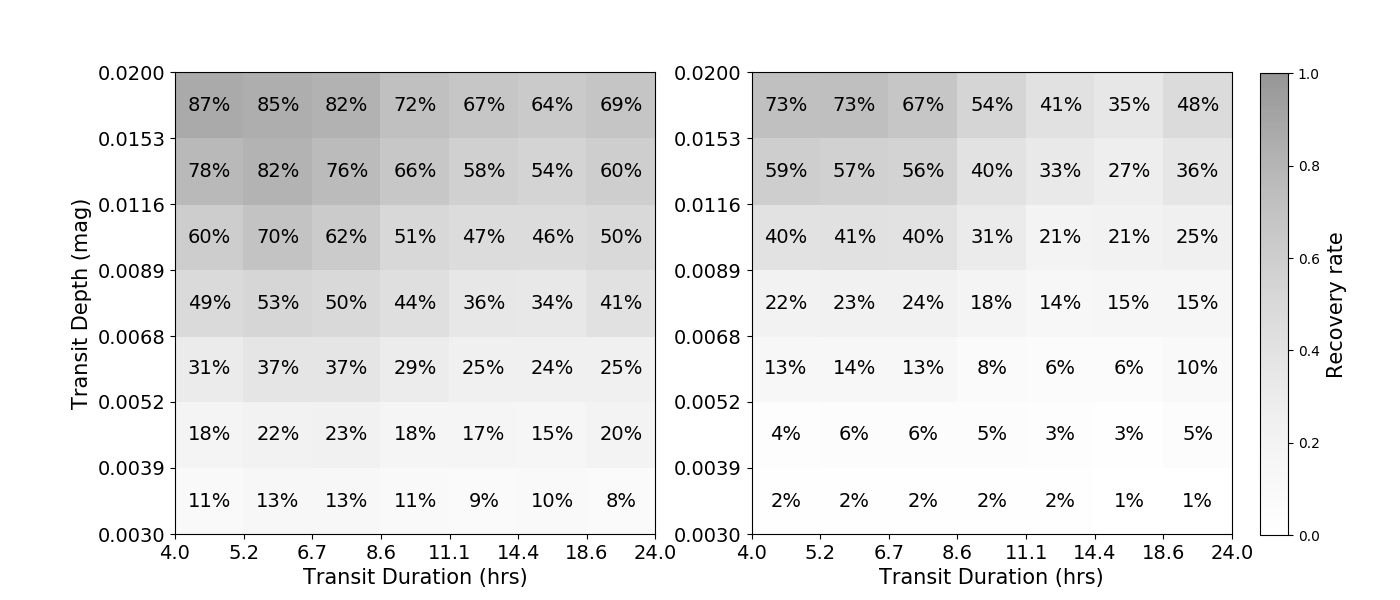}}
\caption{Recovery rates of simulated transits for transit duration-transit depth bins KELT-North (left) and KELT-South (right) light curves. The grayscale bar indicates the fraction of the transits that are correctly recovered, which is also represented by the percent value in each bin.}
\label{fig:rate_duration}
\end{center}
\end{figure}

We display the results of these tests in several ways. The simulated recovery rate distributions across transit depth vs.\ orbital period and transit depth vs.\ transit duration for the KELT-North and KELT-South combined light curves are shown in Figure \ref{fig:rate_period} and Figure \ref{fig:rate_duration}, respectively.


In Figure \ref{fig:rate_period}, the recovery rates decrease with smaller depths and longer periods monotonically. The inserted transit depths and orbital periods are distributed uniformly across the full set of light curves, and so the number of data points in the light curves, stellar density, and photometric uncertainty, are distributed uniformly across Figure \ref{fig:rate_period}.  Therefore from Equation \ref{eqn:SNR_period}, the only parameters that affect the recovery rate difference in this plot are transit depth and orbital period.

\begin{figure}[t]
\begin{center}
\makebox[\textwidth][c]{\includegraphics[width=1.0\textwidth]{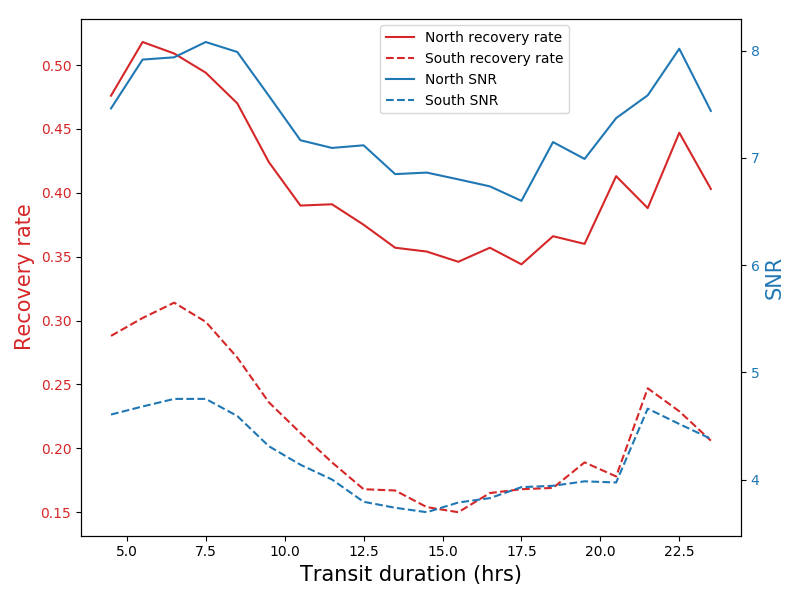}}
\caption{Recovery rate (red line) and calculated SNR (blue line) versus transit duration.  The effects of populating planets evenly in period space, but then converting to transit duration, created a more complex trend in transit duration when accounting for the underlying stellar population.} 
\label{fig:snr_rate}
\end{center}
\end{figure}

Figure \ref{fig:rate_duration} shows the same set of light curves and recovery rates as Figure \ref{fig:rate_period}, but plots transit duration on the horizontal axis instead of Period.  While the transits are distributed uniformly in log-Period space, the transit duration distribution is no longer uniform, because of the varying properties of the stars themselves displayed in Figure \ref{fig:stellar-depth-duration}. In Figure \ref{fig:rate_duration}, the recovery rates no longer vary monotonically from upper left to lower right. This is because stellar densities are no longer uniformly distributed across transit duration, and the stellar density has negative correlation with transit duration (Equation \ref{eqn:duration}).  That is because from Equation \ref{eqn:SNR_duration}, the relation between SNR and transit duration becomes more complex. To examine this effect across transit duration, we divided transit duration into 20 bins. In each bin in transit duration space, we calculated the recovery rate and average SNR, and plot those quantities in Figure \ref{fig:snr_rate}. We see that the overall trends for both SNR and transit recovery are similar in structure across transit duration, indicating that our recovery rates do in fact follow the theoretical expectations.  We can see the reasons for the particular form of that behavior in Figure \ref{fig:stellar-depth-duration}, which shows that at short transit durations the stars are significantly more dense.  That explains the decrease in recovery at very short transit durations.  Conversely, Equation \ref{eqn:SNR_duration} shows that SNR is a stronger function of transit duration than stellar density.  Therefore at long durations, we see an increase in recovery rates.

\begin{figure}[t]
\begin{center}
\makebox[\textwidth][c]{\includegraphics[width=1.0\textwidth]{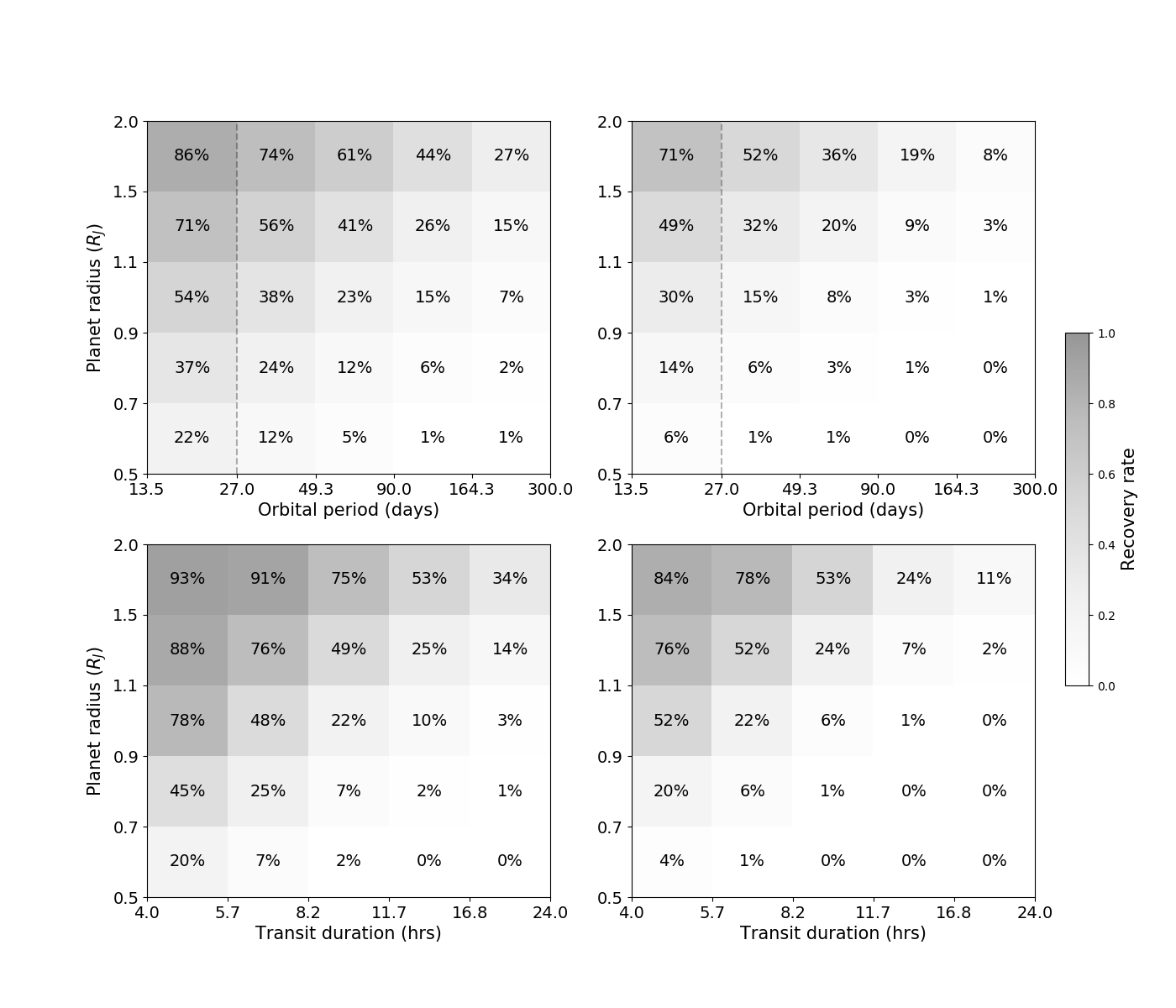}}
\caption{Recovery rate distributions displayed across period vs.\ planetary radius space (top) and transit duration vs.\ planetary radius space (bottom) for the KELT-North (left) and KELT-South (right) combined light curves. The color bar indicates the fraction of the transits that are correctly recovered, which is also represented by the percent value in each bin.}
\label{fig:recovery_by_radius}
\end{center}
\end{figure}

The plots convey the operational efficiency of the existing KELT light curves in terms of the key parameters of the inserted signals: transit depth and orbital period or transit duration.  It is also possible to display the vertical axis not in terms of transit depth, but rather in terms of planet radius $r$, which is shown in Figure \ref{fig:recovery_by_radius}.  The planet radius is calculated based on the inserted transit depth, the host star's radius from the CTL and the limb-darkening coefficients, assuming perfect deblending of the host star from any nearby neighbors.  The listed radius is therefore a ``true" depth rather than an ``observed" depth.  Since the simulated transits were drawn from a logarithmic distribution of depths, rather than radii, a small number of light curves are excluded from the plot, in which the inserted signals represent planets smaller than 0.5 $R_J$.  We have also continued to exclude the small number of initially inserted signals that would represent planets larger than 2 $R_J$.

Because the northern KELT fields have been observed for longer than the southern fields, they have a longer observed time baseline and also more total data points.  Thus the recovery rates shown in Figure  \ref{fig:recovery_by_radius} are greater for KELT-North versus KELT-South.
For the KELT-North light curves, for warm Jupiter systems (planet radius of 0.9 to 1.1 $R_J$ and $13.5 < P < 50$ days) we recover 30\% to 50\% of the transits. For temperate Jupiters ($50 < P < 300$ days), we recover 5\% to 20\% of the transits.  For warm Saturns (planet radius 0.5 to 0.9 $R_J$ and $13.5 < P < 50$ days) we recover 10\% to 30\% of the transits.
The comparable numbers for KELT-South are lower, but not extremely so.

\subsection{Precovery in Practice}
\label{sec:prac}

We can display the results of this analysis in a way that would be more useful for investigators pursuing TESS candidate follow-up.  Importantly, the previous plots considered a successful recovery as one in which the strongest peak in the BLS search corresponded to an orbital period within 0.01\% of the true period.  However, no signal strength criterion was applied.  To characterize signal strength, we use the metric signal-to-pink-noise (SPN).  The SPN ratio is a variant of the SNR computed using a "pink" total noise that includes both uncorrelated (``white") and correlated (``red") noise sources. This ratio is transit model-specific and is calculated as:
\begin{equation}
SPN = \sqrt{ \frac{\delta^2}{\left ( \frac{\sigma_w^2}{N_{pts}} \right ) + \left ( \frac{\sigma_r^2}{N_{transits}} \right )} }    
\end{equation}
where $\sigma_w$ is the white noise estimate, $N_{pts}$ is the number of data points in transit, $\sigma_r$ is the red noise estimate, and $N_{transits}$ is the number of transits observed \citep{Hartman:2016}.
The white noise estimate is effectively the standard deviation of light curve magnitudes. The red noise estimate is obtained in two steps. The light curve is first boxcar-smoothed (binned with overlap) using a width equal to the transit duration. A true ``binned" RMS measurement is obtained from these data. This binned RMS is compared to the expected binned RMS (calculated using the same bin size with the assumption of uncorrelated data) and the red noise component is taken to be the difference (excess). The total pink noise is finally estimated as a combination of the white noise (weighted by the number of points in transit) and the red noise (weighted by the number of transits observed) and compared to the transit depth. By including the a correlated noise component, the SPN tends to provide a significantly fairer estimate of detection significance than would be expected for purely uncorrelated data.

\begin{figure}[t]
\begin{center}
\makebox[\textwidth][c]{\includegraphics[width=1.0\textwidth]{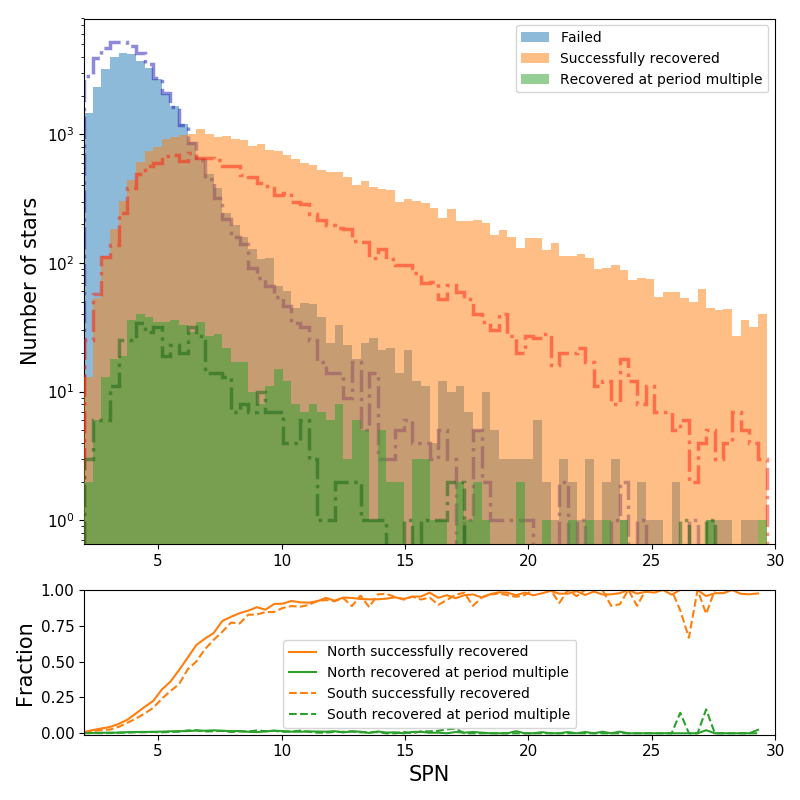}}
\caption{Distribution of SPN values for all light curves, colored by whether the transit signal was recovered with the correct period.  The green color indicates that the recovered period was a low-integer multiple or fraction of the inserted period.  The dot-dashed histogram represents the KELT South case. The lower panel indicates the fraction of the light curves at a given SPN value in which the signal was successfully recovered.  At SPN$\sim6$, half the signals were recovered.}
\label{fig:SPN_rec}
\end{center}
\end{figure}

We now examine the SPN values and how they relate to signal recovery; i.e. how SPN is distributed among the successful versus unsuccessful recovery cases.  That information can be seen for the full set of light curves in Figure \ref{fig:SPN_rec}, showing the distribution of SPN values, along with the fraction of light curves for a given SPN for which the signal was successfully recovered.  At SPN$\sim6$, half the signals were recovered, and at SPN$\sim7$, about three fourths were recovered.

While those fractions provide overall guidance to the SPN threshold to use to determine a likely successful signal detection, it would be more useful to break that number down across the range of observable parameters from the TESS signals.  We therefore compute the distributions from Figure \ref{fig:SPN_rec} across transit depth and transit duration space, and calculate the SPN values for which more than 10\%, 50\% and 90\% of the signals are successfully recovered.

\begin{figure}[t]
\begin{center}
\makebox[\textwidth][c]{\includegraphics[width=1.0\textwidth]{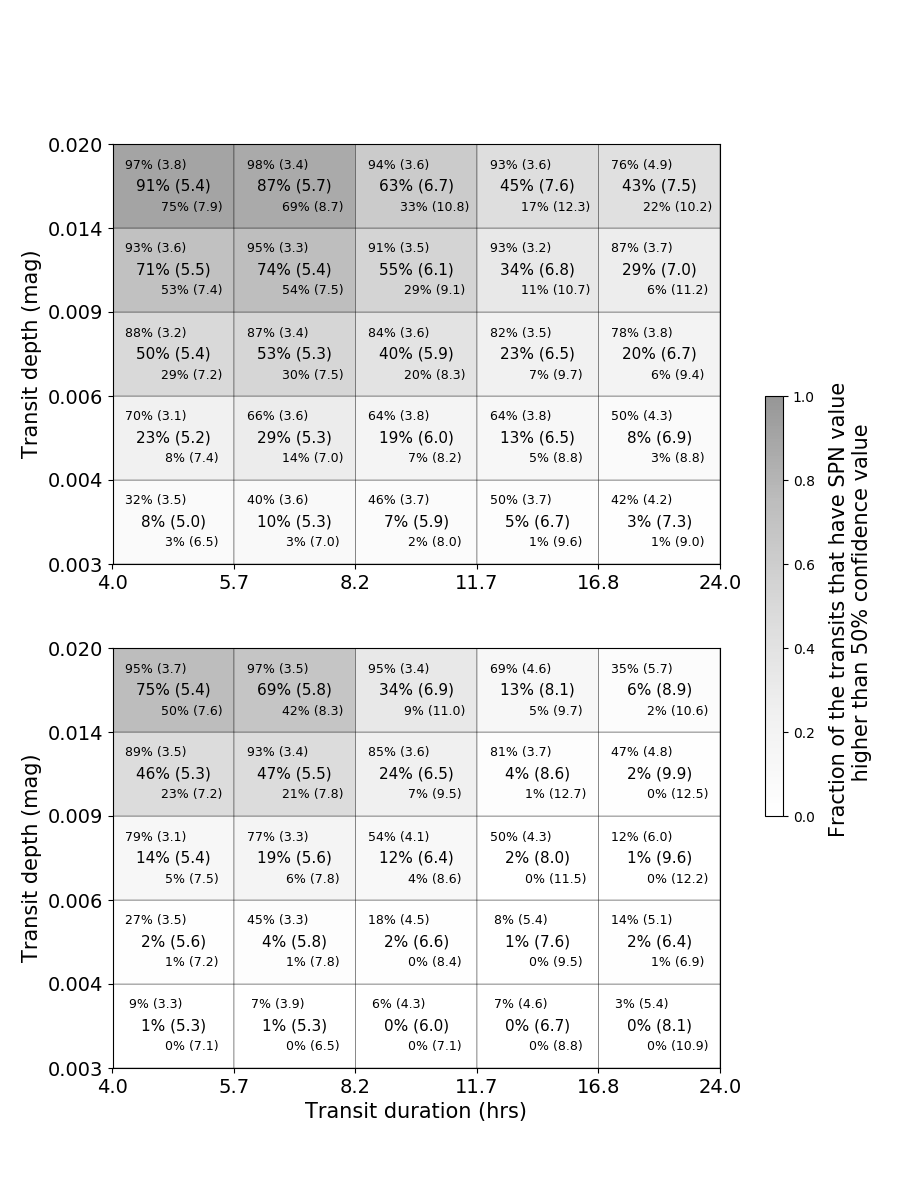}}
\caption{The fractions of KELT-North (top) and KELT-South (bottom) light curves that pass SPN thresholds in transit depth / transit duration bins, and the corresponding SPN values in parentheses. In each box, there are three percentage values and three SPN values.  The percentages reflect the fraction of KELT light curves in that bin of depth/duration space such that if the SPN value is greater than the indicated value, there is a given likelihood that the recovered period is correct.  Those likelihoods are 10\%, 50\%, or 90\%, from upper left to lower right in each bin.  The color bar indicates the fraction of KELT light curves that pass the 50\% confidence threshold.}
\label{fig:spn}
\end{center}
\end{figure}

We display the results of that analysis in Figure \ref{fig:spn}.  The plot displays the confidence that a user can have regarding the ability of the KELT data to detect the signal.  For each bin, we display the fraction of KELT light curves in the bin such that 90\%, 50\%, and 10\% of the light curves above the corresponding SPN limits are recovered at the correct period.  Those numbers provide guidance for how useful the KELT light curves can be for obtaining ephemerides of single-transit signals, depending on the value of individual planet candidates and the availability of observing resources.

For instance, consider the KELT-North bin for signals with transit durations of 8.2 to 11.7 hours and depths of 9 to 14 mmag.  We find that within that bin, if BLS returns an SPN value of 9.1 or greater, there is a 90\% chance that the highest BLS power matches the true period of the inserted transit.  In that bin, 29\% of the KELT light curves yielded SPN values greater than 9.1.  Similarly, 55\% of the light curves in that bin yielded SPN$>6.1$, which is the point where 50\% of all the light curves with SPN larger than that value yielded BLS results where the recovered period matched the inserted period.  In that same way, 91\% of the light curves in that bin yielded SPN$>3.5$, and 10\% of all the light curves with SPN larger than that value yielded BLS results where the recovered period matched the inserted period.

Someone analyzing a single-transit event in TESS with a given transit duration and depth can refer to this figure.  If the target has a high-quality KELT light curve, and the SPN result of a BLS search is greater than the 90\% threshold for that signal, or SPN$>9.1$ for the above example range of depth and duration, they can schedule observations based on the BLS-derived period of that light curve with a 90\% confidence of a successful detection. If observational resources are plentiful, the observer may be willing to accept a 10\% success rate, and can choose to pursue a large number of targets.


For KELT-South, we see that the overall patterns are similar to those for KELT-North, only with lower recovery rates due to the smaller number of data points and shorter time baselines.  With the addition of data that is already acquired, yet currently unreduced for both KELT-North and KELT-South, the recovery rates for both sets should improve. As additional KELT observations are acquired over the next few years, both the time baseline and number of data points will grow, further improving the recovery rates.

Figures \ref{fig:succ_1} through \ref{fig:fail_2} below show example light curves where the inserted signal was successfully or unsuccessfully recovered. A transit with a relatively short period of 38.2 days and shallow depth of 3.5 mmag was successfully recovered in Figure \ref{fig:succ_1}.  A transit with a long period of 255.6 days and deeper depth of 7.8 mmag was successfully recovered in Figure \ref{fig:succ_lp}. In the successful cases, the true period was identified as the strongest peak in the BLS periodogram, as shown in the top panel of Figure \ref{fig:bls_1} and \ref{fig:bls_2}.  The strongest BLS peak in these periodograms has a significant of 4.6$\sigma$ and 3.1$\sigma$ as measured by RMS.  For comparison, we also show the BLS periodograms when not using the TESS information about the transit depth, duration, transit time and ingress/egress time.  In that case, the true period is not the largest BLS peak.

An example of a failure to recover is shown in Figure \ref{fig:fail_1}.  In that case, the transit depth of 4.1 mmag is too shallow compared to the scatter in the KELT data. Figure \ref{fig:fail_2} shows a case where the KELT observations happen to miss the inserted transit completely. The true period is not the largest BLS peak as shown in Figure \ref{fig:bls_3} and \ref{fig:bls_4}.

\begin{figure}[t]
\begin{center}
\makebox[\textwidth][c]{\includegraphics[width=1.0\textwidth]{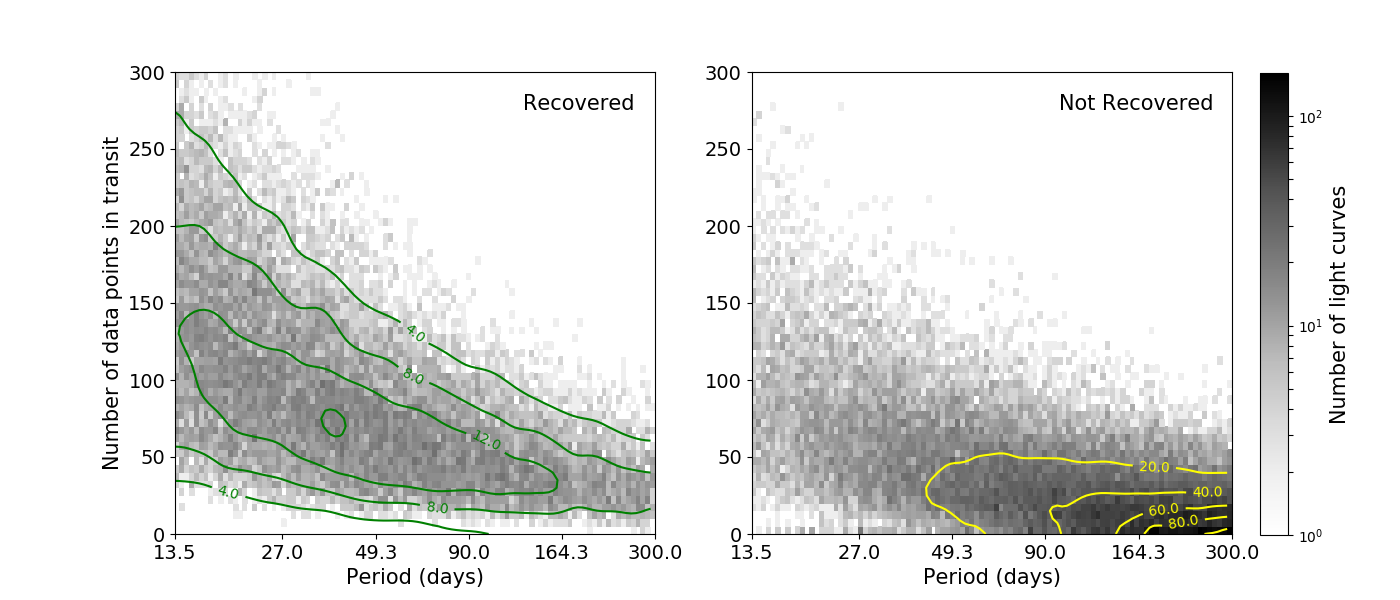}}
\caption{Heatmap with contours showing the number of in-transit data points vs. orbital period in all recovered cases (left) and unrecovered cases (right). The colorbar and contours indicate the number of light curves.}
\label{fig:nit}
\end{center}
\end{figure}

It is useful to investigate how many inserted transits failed to be recovered due to missing data compared to insufficient SNR.  Figure \ref{fig:nit} shows the number of in-transit data points in recovered and unrecovered cases.  We find that the unrecovered cases certainly have fewer points in transit overall.  We do not see any obvious substructure in the distribution of successful recoveries in period space.  That is, we do not see signs of a transit window function, aside from the monotonic decrease in transit recovery with longer orbital periods.

\subsection{Considerations for 2-min Targets vs. FFIs}

\begin{figure}[t]
\begin{center}
\makebox[\textwidth][c]{\includegraphics[width=1.0\textwidth]{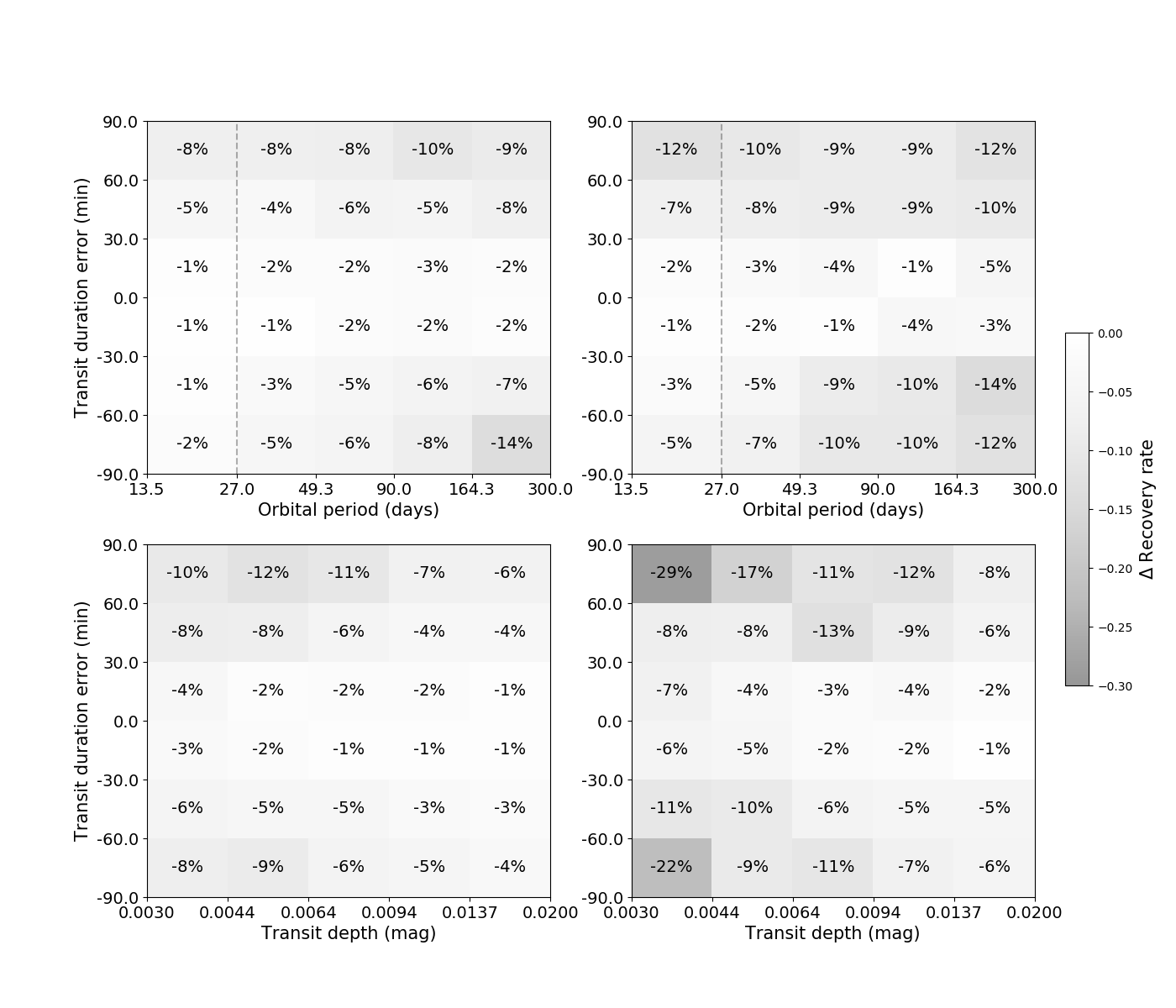}}
\caption{Recovery rate error distribution exploring the effect of an error in the TESS inserted transit duration.  The plot shows the recovery rate for a range of duration errors versus orbital period and transit depth for the $\sim$29,000 KELT-North light curves (left) and $\sim$14,000 KELT-South light curves (right) with simulated transits for which the transits were successfully recovered.  These plots represent the degree to which to recovery rates degrade compared to Figure \ref{fig:rate_period}.}
\label{fig:durerr}
\end{center}
\end{figure}

\begin{figure}[t]
\begin{center}
\makebox[\textwidth][c]{\includegraphics[width=1.0\textwidth]{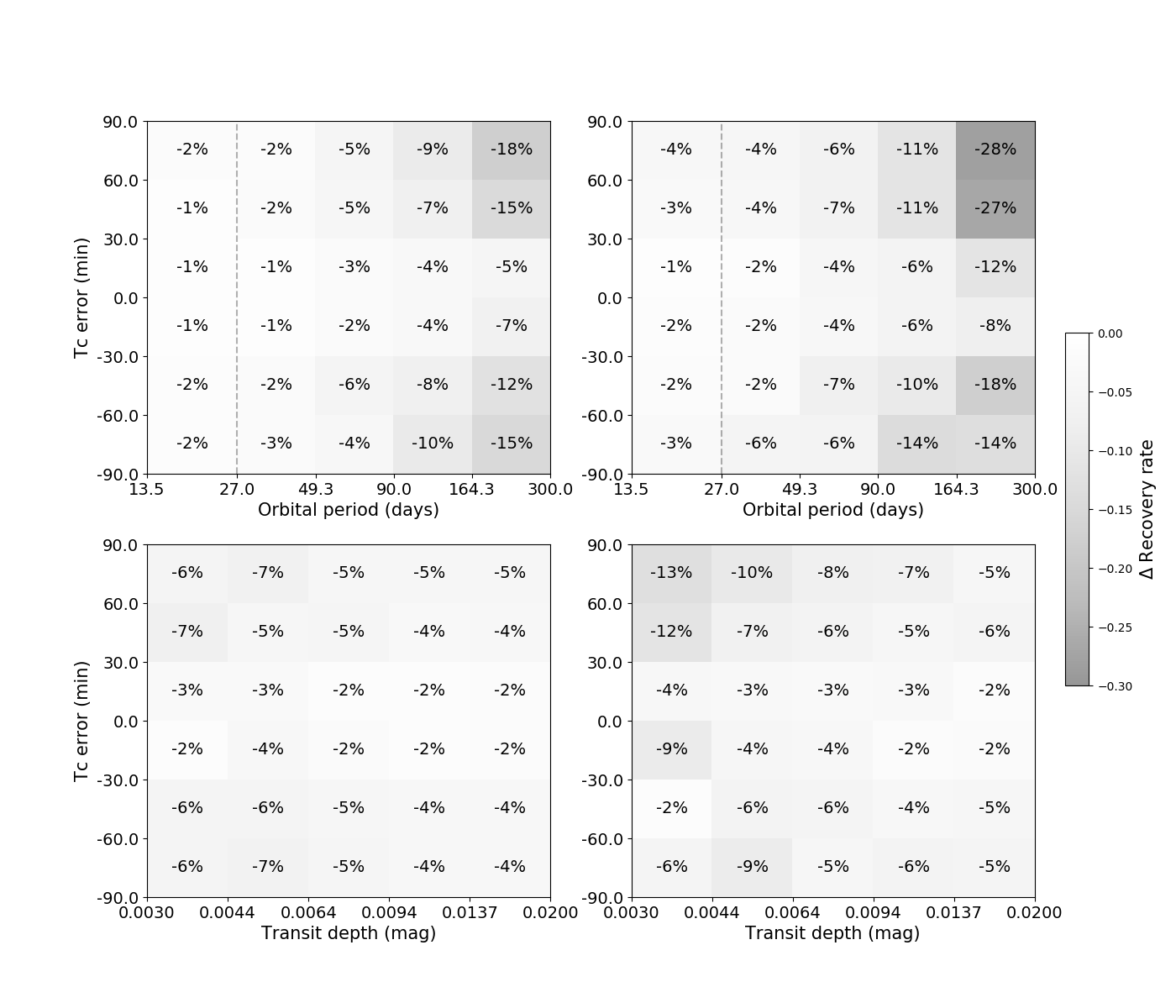}}
\caption{Recovery rate error distribution exploring the effect of an error in the TESS inserted transit time $T_c$. The plot shows the recovery rate for a range of $T_C$ error versus orbital period and transit depth for the $\sim$29,000 KELT-North light curves (left) and  $\sim$14,000 KELT-South light curves (right) with simulated transits for which the transits were successfully recovered in the previous step.  These plots also represent the degree to which to recovery rates degrade compared to Figure \ref{fig:rate_period}.}
\label{fig:$T_C$err}
\end{center}
\end{figure}

The tests described so far do not depend on whether the TESS planet candidates are extracted from the 2-min postage stamps or the 30-min FFIs.  The main difference between the two types of data that we consider here are the longer cadence of the FFIs, which could lead to greater errors in the measured transit duration of the resulting single transits.  We examine that effect by taking a subsample of the successfully recovered light curves and add a random error to the transit duration and transit time $T_C$ ranging from -90 to +90 minutes, respectively.  Figures \ref{fig:durerr} and \ref{fig:$T_C$err} shows the impact of the duration error and $T_C$ error on the recovery rates for different period and depth regimes in KELT North and South data. We find that any uncertainty in the duration and $T_C$ caused by the lower cadence of the 30-minute FFIs causes at most a $\sim$15\% reduction in recovery rates.

\section{Discussion} 
\label{sec:disc}

Since this work demonstrates that KELT and similar ground-based surveys have the ability to detect significant numbers of long period exoplanets, it is natural to ask why the surveys have not already detected them.  That is, the simulations described in this paper do not use actual TESS light curves, rather just signals inserted into archival KELT lightcurves, so if such signals are already in there, why are they not already found?  In fact, none of the three most productive ground-based transit surveys (WASP, HAT, and KELT) have detected a transiting planet with an orbital period longer than 20 days.

The reason such signals have not been discovered so far is that long-period planets have smaller SNR, especially for transit surveys that search in phase-folded data, along with the fact that at longer periods the transit probability is intrinsically smaller.  Therefore ground-based surveys would have to intensively look for rare, low-SNR signals, which usually means lowering a SNR threshold on an automated search algorithm.  That will result in a large number of spurious false positive candidates, which themselves are more difficult to follow up and confirm for long-period and long-duration signals.  Existing surveys like KELT simply do not have the resources to chase that many candidates.

However, the availability of the TESS signal provides additional information that changes the situation.  Because TESS sees a transit at a particular time, for a particular star, with a particular transit duration, the parameter space that must be searched by an algorithm like BLS shrinks dramatically.  For a given light curve, it is no longer necessary to search through a range of transit phases and transit durations.  Rather, the $T_C$ is known, along with the duration, and the only unknown parameter is the period.  Therefore the SNR threshold is effectively lowered, and smaller signals can be detected since the spurious signal rate is lowered.  Mostly, though, the availability of the TESS signals means that the particular stars with signals are known, shrinking the number of light curves to be searched from hundreds of thousands to hundreds, since only those stars whose light curves that show a signal in TESS need to be searched.  That means that the total number of candidates that emerges from the automated cuts is reduced by several orders of magnitude, and can be followed up with a reasonable set of observing resources.

\subsection{Potential for Improvement}

The procedures outlined here represent a first attempt to quantify the value of ground-based transit surveys for detecting single-transit events in TESS.  There are a number of steps in which we have simplified the analysis, and which can be improved in future work.  Some of those steps involve a more realistic set of planet-star system properties.  Although we are using real KELT light curves and the actual stellar properties for those stars in from the TIC, we have assumed circular and equatorial transits.  A realistic distribution of those parameters would allow for a more accurate set of likely recovery rates.  It would also be useful to redo the analysis using realistic transit models that account for limb-darkening, rather then using box-shaped inserted transits.  In general, we would not expect that difference to have a large effect \citep{Gould:2006}, but it should be investigated especially for stars observed with the 30-minutes cadence by TESS.  We can also explore how transit timing variations (TTVs) would affect the recovery rates.  We would not expect a large fraction of these planets to have TTVs at the level that would interfere with our detection methods \citep{Holczer:2016}, but that is worth investigating.

Furthermore, the analysis in this paper has addressed the overall sensitivity of the process - we have not attempted to perform an actual simulation of the total numbers of single-transit event TESS will detect, and the resulting total numbers that can be precovered using the KELT data.  We intend to conduct such an exercise, using the \citet{Villanueva:2018} simulations.

A key issue we have not addressed here is the combination of existing ground-based photometry and radial velocity (RV) measurements.  A small fraction (about 5\%) of the cases included in this analysis are systems where the inserted transit signal was found at a fractional multiple of the true period.  Without additional information, those cases do not represent strong opportunities for follow-up, since it would be necessary to conduct several additional observations to investigate each separate multiple of the precovered period. However, if RV observations were available that ruled out one or more of the common multiples, the probability of catching the transit would significantly increase.  In fact, RV observations could enhance the value of photometric precovery overall by constraining the possible orbital periods.  The observational needs for such an effort is described in \citet{Villanueva:2018} and also in Dragomir et al. (in preparation).  The availability of photometric precovery should be part of such a project, allowing observers to conserve resources by maximizing the probability of ephemeris conformation.

In a similar way, the assumption of a circular or near-circular orbit can be used to estimate the orbital period as outlined by \citet{Yee:2008}.  That information can be used to constrain the precovery results to improve the reliability of the precovery-derived periods at a given SNR, or to increase the number of candidates that can be investigated by selecting a lower SNR cutoff at the same confidence of correct transit recovery.

One final point is that while KELT might be the best-positioned of the current ground-based transit surveys for TESS single-transit precovery, other survey data should also be explored.  The SuperWASP, HATNet, and HAT-South surveys should have the ability to conduct a similar sort of precovery, and do even better for the lower-mass stars to which KELT is less sensitive.  That is especially the case for the recently started Next Generation Transit Survey \citep[NGTS]{Wheatley:2018}, and the Multi-site All-Sky CAmeRA (MASCARA) survey \citep{Talens:2017} should be particularly sensitive to transits of very bright ($V<8$) stars.  Combining data for the same star from multiple surveys with complementary time baselines and observing seasons offers an especially rich possibility for transit detection, as noted by \citet{Fleming:2008}.

In fact, the approach described in this paper could just as well be applied to other space-based missions, such as CoRoT, Kepler, and K2.  For CoRoT and Kepler, those missions observed for a long enough time baseline that any single-transit candidates from those missions would have true periods much longer than could be identified by the wide-field ground-based surveys like KELT, SuperWASP or HAT.  Furthermore, most of the Kepler targets are significantly fainter than the optimal magnitude range of the ground-based surveys. K2 presents an interesting opportunity, since the observing campaigns last about 80 days, and the typical targets are a lot brighter than for the original Kepler field.  Unfortunately, KELT mostly avoids the ecliptic, missing most of the K2 fields, and we have not found any single-transit K2 candidates in the areas with KELT overlap.  Other ground-based surveys may be better positioned for transit precovery with K2.

\subsection{Planet Confirmation}

Precise ephemerides are important for determining the planetary nature of TESS transit candidates, whether via dynamical confirmation or statistical validation \citep{Morton:2012,Torres:2011}.  Reliable ephemerides allow radial-velocity observers to schedule observations at quadrature for efficient dynamical confirmation of planets.  However, the ephemerides are especially crucial for efficiently planning photometric observations to conserve observing resources.  Because the TESS pixels are very large on the sky (about 20 arcsec across) a large fraction of TESS transit candidates will be blended with nearby stars in the TESS photometry.  That will create uncertainty about which star is experiencing the transit, diluting the depth of the transit signal itself, and thus affecting the derived planet radius.  Depending on the degree of blending and the depth of the transit, the TESS data, along with high-resolution images of the region around the target, can be used to determine the target star and true transit depth.  But that will not be the case in all circumstances, primarily for crowded regions and shallow transits.  It will therefore be necessary to conduct additional time-series photometry with higher spatial resolution to determine whether the signal is coming from the expected target star or a nearby blended eclipsing binary.  Thus, the large TESS pixels lead to a greater reliance on photometric follow-up for planet confirmation compared to a survey with smaller pixels, and the increased importance of photometric follow-up leads to a corresponding importance on reliable ephemerides.

The TESS mission is planning follow-up observations to confirm planet discoveries through the TESS Follow-On Program (TFOP).  TFOP will be focused on confirming the primary targets of the TESS mission, which are small planets with radii smaller than 4 $R_{\Earth}$.  There is also a project being developed at the NASA Exoplanet Science Institute (NExScI) called ExoFOP-TESS to organize follow-up observations for all other TESS transit candidates, including giant planets and candidates detected in the FFIs.  Given the sensitivity of the KELT photometry, this kind of precovery will be particularly useful for the ExoFOP-TESS project.


\section{Conclusion}

Between the targeted postage stamps and the FFIs, the TESS data will contain tens of thousands of detected transiting exoplanets, amidst hundreds of thousands of false positive signals, most of which will be eclipsing binary stars \citep{Sullivan:2015}.  Having precise and reliable ephemerides for the TESS transit candidates will be crucial for two reasons: planning confirmation observations, and conducting follow-up characterization studies. The recent announcement of the launch delay for JWST until 2021 means that even more time will pass between the TESS observations and JWST atmospheric observations, exacerbating the effects of imprecision of ephemerides as seen by \citet{Benneke:2017}.

This work demonstrates that the use of ground-based photometric data can be used to recover the ephemerides of TESS single transit events.  Specifically, we find that when given information about a single transit from TESS data ($T_C$ and transit duration), the signal of the phase-folded transit can be detected in KELT data, and thus determine the ephemerides, of 30\% to 50\% of warm Jupiters, of 5\% to 18\% of temperate Jupiters, and 10\% to 30\% of warm Saturns and Neptunes.  This technique can likewise be applied to data from other ground-based transit surveys in conjunction with TESS, further increasing the number of single-transit events from TESS with precisely determined ephemerides.  

The work described here represents a proof of concept for the approach.  In future papers, we intend to explore how the efficacy of precovery methods like this depends on other system parameters, such as the host star properties.  We will also explore the particular value of this approach in cases where TESS sees two or three transits.  In such cases, the initial ephemerides will be much more constrained by the TESS data directly, and we will examine the marginal increase in precision provided by data from ground-based data, rather than the overall recoverability question we probe in this paper.

\acknowledgments{
This research made use of NASA’s Astrophysics Data System, and the SIMBAD database, operated at CDS, Strasbourg, France.  This research has also made use of the NASA Exoplanet Archive, which is operated by the California Institute of Technology, under contract with the National Aeronautics and Space Administration under the Exoplanet Exploration Program. Work by S.V.Jr. is supported by the David G. Price Fellowship for Astronomical Instrumentation and by the National Science Foundation Graduate Research Fellowship under Grant No. DGE-1343012.  Work by J.P. and X.Y. is supported by the NASA K2 Guest Observer Cycle 5 Award 80NSSC18K0300 under solicitation NNH16ZDA001N.
}

\bibliographystyle{apj}
\bibliography{main.bbl}

\begin{figure}[t]
\begin{center}
\makebox[\textwidth][c]{\includegraphics[width=1.0\textwidth]{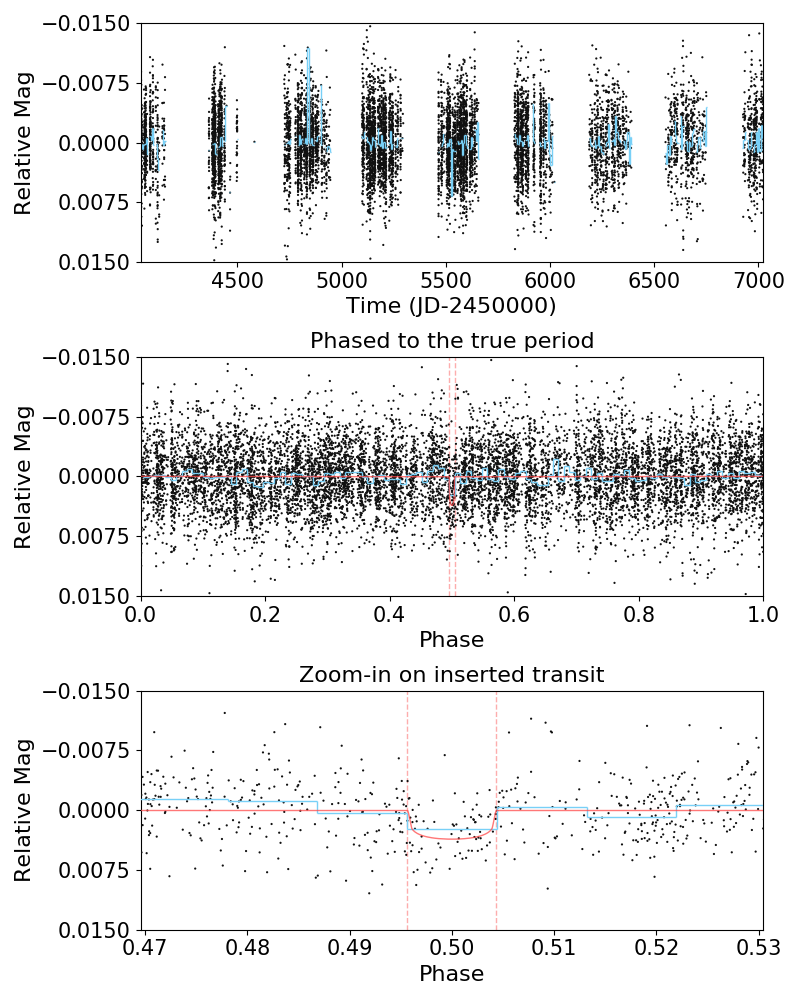}}
\caption{An example of a successfully recovered transit. The top panel shows the unphased light curve with the transit inserted; the middle panel shows the light curve phased at the true period, and the bottom panel shows the zoom-in on the inserted transit. The blue line represents a median bin of the phased data, and the red line represents the inserted transit model. The period of the transit is 38.2 days, with a duration of 7.6 hrs and a depth of 3.5 mmag.  The light curve has an RMS when binned at the duration of the transit of 0.7 mmag.}
\label{fig:succ_1}
\end{center}
\end{figure}

\begin{figure}[t]
\begin{center}
\makebox[\textwidth][c]{\includegraphics[width=1.0\textwidth]{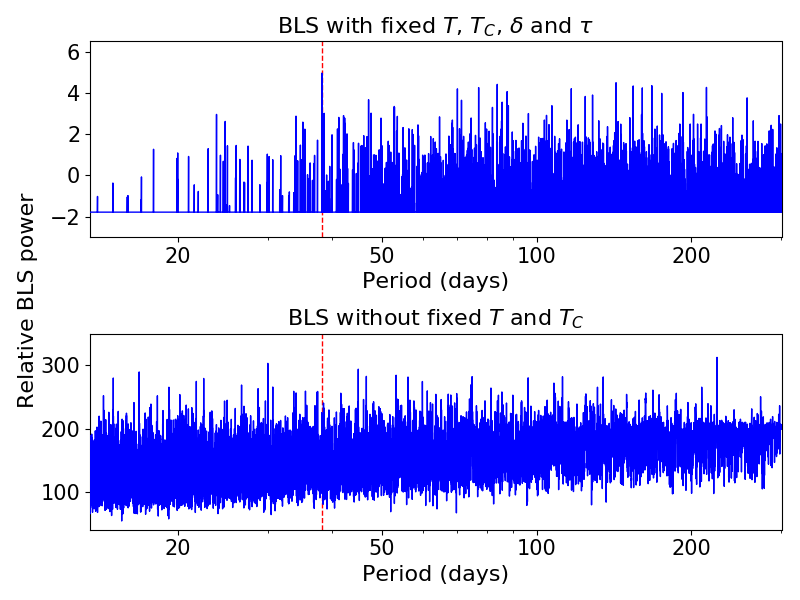}}
\caption{The BLS periodogram for the successfully recovered transit shown in Fig. \ref{fig:succ_1}. The top panel shows the BLS power distribution for BLSFixDurTc with fixed depth, transit duration, transit time and duration of ingress, and the bottom panel shows the case for a BLS search without a fixed ephemeris. The period corresponding to the peak BLS power in the top panel is the same as the inserted period (the red dashed line), while the peak BLS power in the bottom panel is different from the inserted period.}
\label{fig:bls_1}
\end{center}
\end{figure}

\begin{figure}[t]
\begin{center}
\makebox[\textwidth][c]{\includegraphics[width=1.0\textwidth]{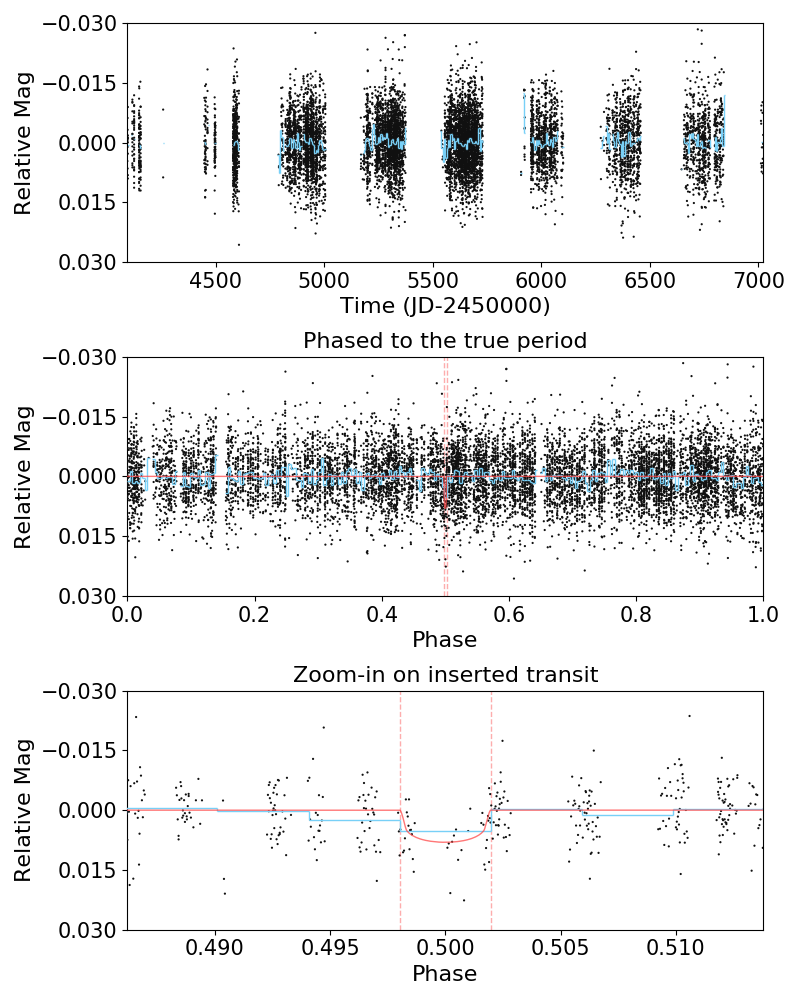}}
\caption{Another example of a successfully recovered transit, similar to figure \ref{fig:succ_1}. The BLS output period is 255.6 days, with a duration of 22.5 hrs and a depth of 7.8 mmag.  The light curve has an RMS when binned at the duration of the transit of 1.6 mmag.}
\label{fig:succ_lp}
\end{center}
\end{figure}

\begin{figure}[t]
\begin{center}
\makebox[\textwidth][c]{\includegraphics[width=1.0\textwidth]{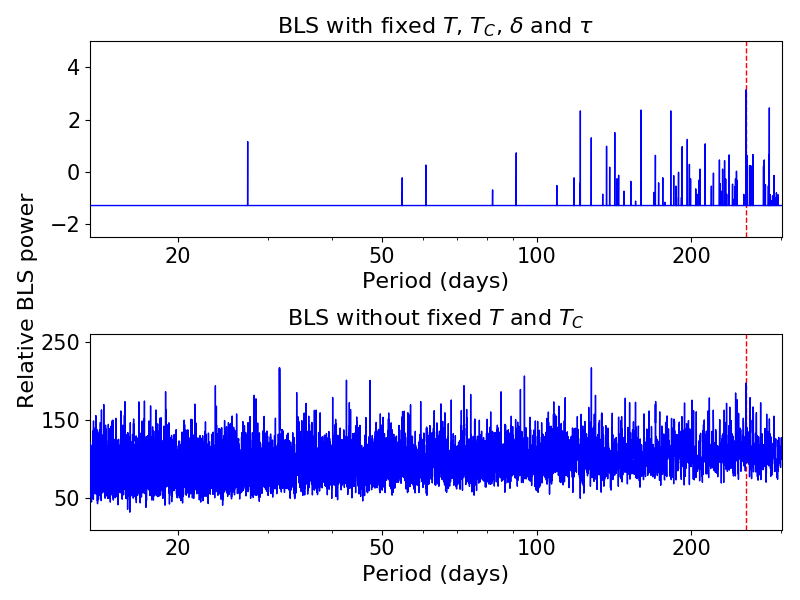}}
\caption{The BLS periodogram for the successfully recovered transit shown in Fig. \ref{fig:succ_lp}. The top panel shows the BLS power distribution for BLSFixDurTc with fixed depth, transit duration, transit time and duration of ingress, and the bottom panel shows the case for a BLS search without a fixed ephemeris. The period corresponding to the peak BLS power in the top panel is the same as the inserted period (the red dashed line), while the peak BLS power in the bottom panel is different from the inserted period.}
\label{fig:bls_2}
\end{center}
\end{figure}

\begin{figure}[t]
\begin{center}
\makebox[\textwidth][c]{\includegraphics[width=1.0\textwidth]{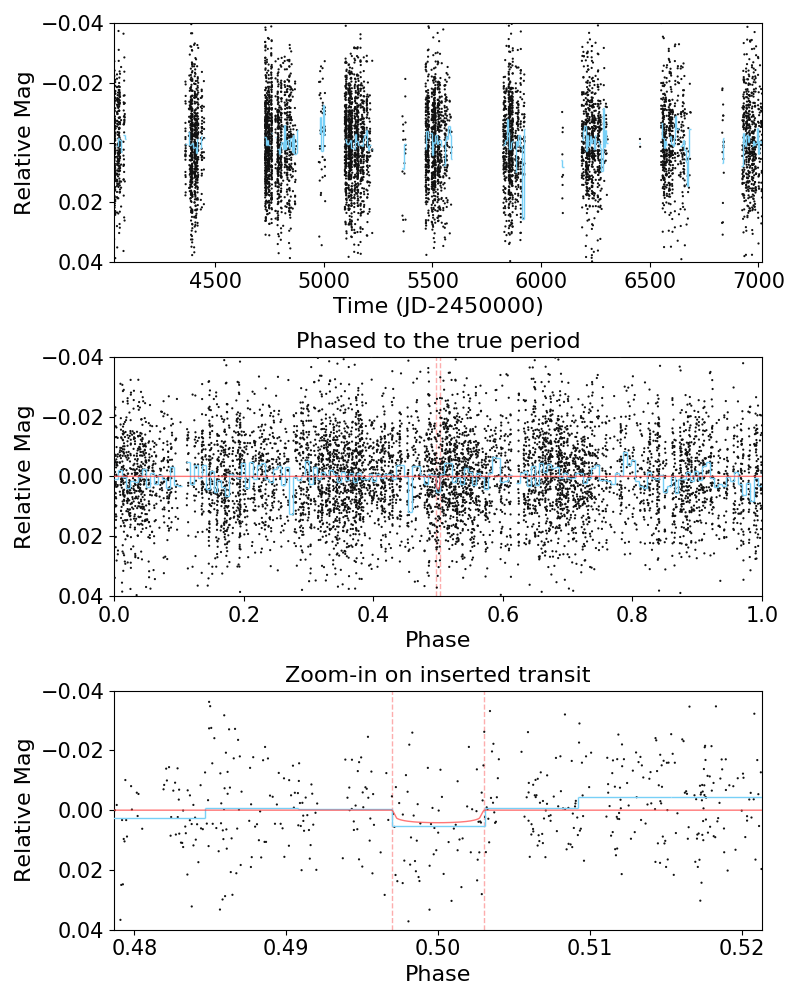}}
\caption{An example of a failed transit recovery. In this case, the attempted recovery period from the BLS analysis is 16.7 days, while the inserted period is 84.9 days. The inserted transit duration is 11.8 hrs, and the depth is 4.1 mmag.  The light curve has an RMS when binned at the duration of the transit of 3.1 mmag.}
\label{fig:fail_1}
\end{center}
\end{figure}

\begin{figure}[t]
\begin{center}
\makebox[\textwidth][c]{\includegraphics[width=1.0\textwidth]{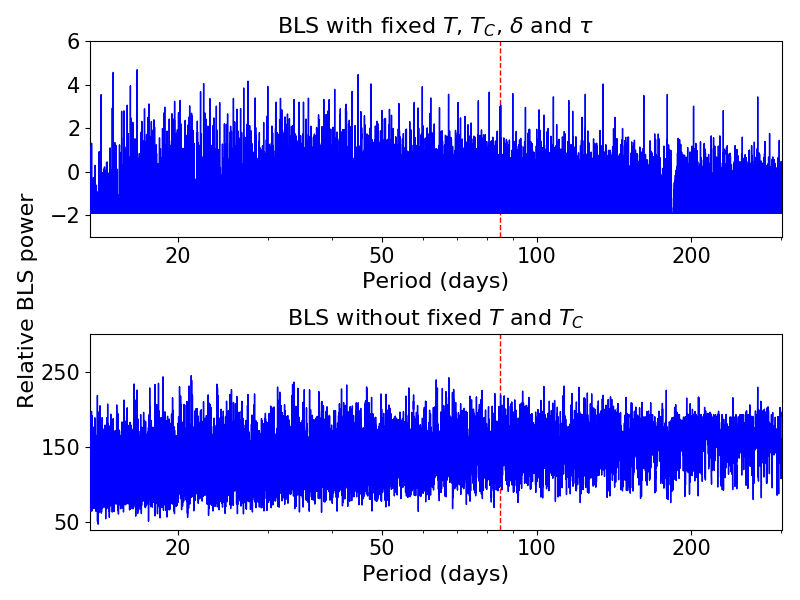}}
\caption{The BLS periodogram for the failed recovered transit shown in Fig. \ref{fig:fail_1}. The peak BLS power in both panels are different from the inserted period (the red dashed line).}
\label{fig:bls_3}
\end{center}
\end{figure}

\begin{figure}[t]
\begin{center}
\makebox[\textwidth][c]{\includegraphics[width=1.0\textwidth]{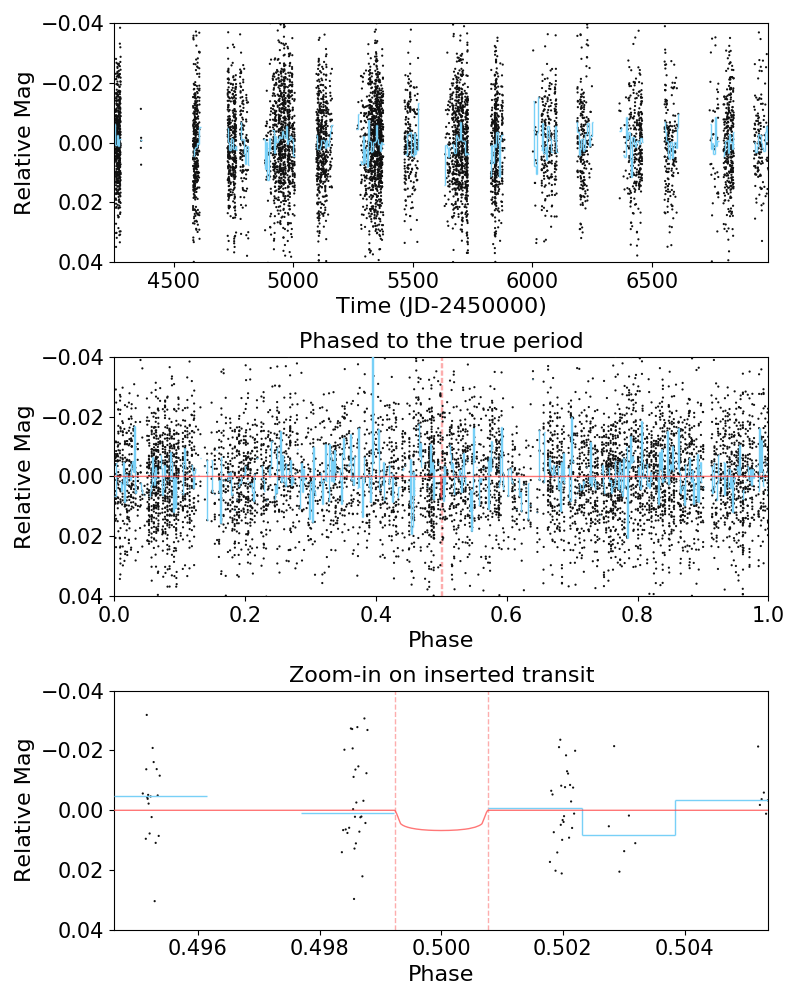}}
\caption{Another example of a failed transit recovery. In this case the KELT data missed the event. The attempted recovery period from the BLS analysis is 66.8 days, while the inserted period is 292.0 days, with a duration of 10.1 hrs and a depth of 6.6 mmag.  The light curve has an RMS when binned at the duration of the transit of 6.7 mmag.}
\label{fig:fail_2}
\end{center}
\end{figure}

\begin{figure}[t]
\begin{center}
\makebox[\textwidth][c]{\includegraphics[width=1.0\textwidth]{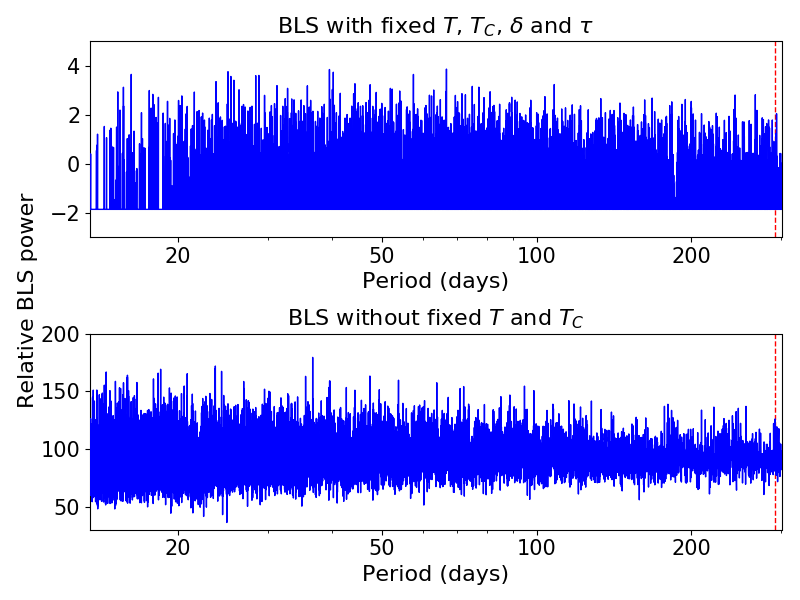}}
\caption{The BLS periodogram for the failed recovered transit shown in Fig. \ref{fig:fail_2}. The peak BLS power in both panels are different from the inserted period (the red dashed line).}
\label{fig:bls_4}
\end{center}
\end{figure}

\end{document}